\documentclass[useAMS,usenatbib]{mn2e}
\usepackage{graphicx,epstopdf}

\title{Revisiting two local constraints of the Galactic chemical evolution}
\author[M. Haywood]
       {M. Haywood\thanks{email : Misha.Haywood@obspm.fr} \\
        GEPI, Observatoire de  Paris, F-92195 Meudon Cedex,France}
\date{Accepted.
      Received ;
      in original form }
\pagerange{\pageref{firstpage}--\pageref{lastpage}}
\pubyear{2006}          
\begin{document}         

\maketitle
\begin{abstract}  
I  review the  uncertainties  in two  observational
local  constraints  of the  Galactic  disc  chemical  evolution :  the
metallicity distribution of  long-lived dwarfs and the age-metallicity
relation.   
Analysing most recent data, it is shown first that the observed metallicity distribution
at solar galactocentric radius, designed with standard methods, is more fit 
to a closed-box model than to the infall metallicity distribution.
We argue that this is due to the specific contribution of the
thick disc population, which has been overlooked in both the derivation of the observed metallicity distribution
and in standard chemical evolution models. 
Although this agreement disqualifies the metallicity distribution as the best supportive 
(indirect) evidence for infall, we argue that the evolution must be more complex than described by 
either the closed-box or standard infall models.

It is then shown that recent determinations of the age-metallicity distribution from large Str\"omgren photometric
surveys  are dominated by  noise resulting  from systematic  biases in
metallicities  and  effective  temperatures.   These
biases are  evaluated and a
new age-metallicity distribution  is obtained, where particularities
of  the previous determinations  are phased out. The
new age-metallicity relation shows a mean increase limited to about a factor of 2 in Z
over the disc age.
It is shown  that below 3 Gyrs, the
dispersion in  metallicity is about  0.1 dex, which, given  the  
observational uncertainties  in  the  derived
metallicities, is compatible with the small cosmic dispersion measured
on the ISM and meteoritic  presolar dust grains.  A population that is
progressively  older  and  more  metal-rich arises  at  a  metallicity
greater  than that of the  Hyades, to  reach [Fe/H]$\approx$+0.5  dex  at ages
greater  than  5  Gyrs.  We suggest that this is best explained by radial
migration.  A symmetrical  widening of  the metallicity
interval towards lower values is seen at about the same age, which is
attributed to a similar  cause. 
Finally,  
the  new  derived  ages  are  sufficiently  consistent  that  an
age-metallicity relation  within the  thick disc is  confirmed.  These
new features  altogether draw a  picture of the chemical  evolution in
the solar neighbourhood where  dynamical effects and complexity in the
age-metallicity distribution dominate,  rather than a generalised high
dispersion at all ages.
\end{abstract}

 
\section{Introduction} 
The metallicity distribution of long-lived dwarfs and the age-metallicity relation 
are two often-quoted constraints of the chemical evolution of the Galactic disc: 
the latter constraint is considered to be weak, whereas the former one is considered to be a strong, or even the 'strongest' constraint.
Both conclusions are usually perceived as definitive\footnote{In a recent article on modeling of the 
chemical evolution of the Milky Way, the authors (Romano et al. 2005.) argue that
'the uncertainties in the data have become really small'
to focus on theoretical uncertainties. A similar conclusion is reached by Nordstr\"om et al. (2004)
on the 'G-dwarf problem'.} observational results.
On the first point,  this is illustrated by
the vast majority of papers published in the last decade on the chemical evolution of the Galactic disc, which 
is described  similarly with prolonged infall controlling its progressive 
inside-out formation. 
The main  motivation for adopting infall being the capacity to reduce the number of 
metal poor stars born at early ages, providing a solution to the so-called 'G-dwarf' problem.
How robust is the observational basis - the dwarf metallicity distribution -  
of such uniformity in models ? 
The unbiased metallicity distribution of dwarfs is usually obtained by 
weighting the metallicity bins of local samples using
a relation between metallicity and  the vertical velocity dispersion ($\sigma_W$) as 
measured on these same samples. 
Due to the small statistics in the lower metallicity bins and the uncertainties that plague
the photometric metallicity scales, this process leads to large 
uncertainties in the level of corrections that are applied, as can be testified by the variety of
[Fe/H]-$\sigma_W$ relations found in the literature (see Fig. 3).
The recent release of large catalogues of solar neighbourhood stars (Nordstr\"om et al. 2004, Valenti \& 
Fischer 2005) containing radial velocities should significantly improve the situation.
The level that this relation should reach at [Fe/H]$<$-0.4 dex is usually not questioned (with
the exception of Sommer-Larsen (1991)), however it is shown here that the inclusion 
of the thick disc stars introduces significant
modifications to the diagnostic of the 'G dwarf problem'.
After having questioned the metallicity scales in Haywood (2001, 2002), 
we point here to the ambiguity caused by the thick disc population, which, 
if included in 
local samples, makes the observed distribution compliant with the predictions of the 
simple closed-box model.
This suggests that some caution should be 
exercised as long as most of the  evidence for prolonged infall is limited to the  
so-called 'G dwarf problem'.

Since Edvardsson et al. (1993) the age-metallicity distribution (hereafter AMD) is often 
commented as being highly dispersed at all ages, providing a poor constraint to 
chemical evolution models of the Galaxy. Two recent studies, based on large photometric datasets, have claimed
a similar result (Feltzing, Holmberg \& Hurley 2001; 
Nordstr\"om et al. 2004).
However, and though it has received little attention
(but see Pont \& Eyer 2004), these recent AMDs illustrate a rather  atypical chemical evolution, 
where the youngest  stars ($<$3 Gyrs) show the largest range 
of metallicity,  stars between 5 and 10 Gyrs show negative evolution 
(i.e. mean decreasing metallicity with age), and the lowest metallicity  stars
([Fe/H]$<$-0.5) are more often 'young' (age $<$ 5 Gyrs) 
than old. Although such non-conventional evolution is not
excluded (one can think of an enhanced infall episode increasing the dilution of metals in the
ISM), confirmation is needed. Moreover, the large dispersion at all ages is difficult to reconcile 
with the very small dispersion that is measured either in the local ISM (Cartledge et al. 2006), on meteoritic 
pre-solar grains (Nittler 2005), or abundance ratios, which all suggest that the gas from which stars are born is very 
homogenenous at all times in the disc, the thick disc, and probably the halo (Arnone et al. 2005)

In the following section, we present a discussion of recent metallicity distributions 
and scale height corrections, while the third section focuses on the biases in  the 
age-metallicity relation. Section 4 presents a derivation of a corrected AMD, which although not 
optimised, is quite different from the previously cited studies. 
We conclude in section 5.

\section{The metallicity distribution}

We first review the metallicity scales used in recently published studies of the 
local metallicity distribution, then focus on the scale height corrections.

\subsection{Photometric metallicity scales}

In discussing the local metallicity distribution, the value at which this distribution peaks
is a key issue, related to the thick disc. 
In a closed-box model centred on solar metallicity, the percentage 
of material which forms between [Fe/H]=-1.0 and -0.50 is about 18-20\%. 
Allowing for a thick disc with about 5-8\% of the local stellar density and a
scale height of about 800-1000 pc, we arrive at a percentage of 16-21\% of the total 
stellar surface density (assuming 250 pc scale height for the thin disc).
Given the uncertainties in these parameters, the 2 estimates can be 
considered as compatible.
If the observed peak was at [Fe/H]=-0.2 dex or lower, the predicted percentage of stars between 
-1.0$<$[Fe/H]$<$-0.5 in a closed box model centred on this value (-0.2 dex) would be higher 
than $\approx$ 30\%, a value incompatible with the thick disc characteristics. 
The thick disc and photometric metallicity scales are therefore at the centre of the discussion about 
the mean metallicity of the solar neighbourhood stars, but the literature 
published on the subject since 1989 shows this mean has a high variability. The 
mode of the metallicity distribution of long-lived dwarfs in the solar 
neighbourhood has fluctuated between -0.4 dex (Pagel 1989) and -0.05 dex 
(Haywood 2001), with intermediate values by Wyse \& Gilmore (1995), 
Rocha-Pinto et al. (1996), \cite{02KOT_EA2} and J{\o}rgensen (2000).  
This is somewhat surprising since, for example, over the same period, the same
quantity for halo stars has been left relatively undisputed at $\approx$ -1.5 dex.
One possible reason is that halo stars are often measured spectroscopically, while, 
paradoxically, systematic spectroscopic surveys of nearby disc stars have commenced only very recently 
(Luck \& Heiter 2005; Valenti \& Fischer 2005; Allende Prieto et al. 2004), impulsed
in particular by planet-search programs.

Since Haywood (2001), a few studies relevant to the 'G-dwarf problem' have been published, 
in particular from \cite{02KOT_EA2}, and the spectroscopic complete surveys 
just mentioned. The metallicity scale by \cite{02KOT_EA2} differs from ours
by a significant amount (0.2 dex), and we go here into some detail explaining the origin of this
difference. Then we compare with spectroscopic surveys. 

\subsubsection{Kotoneva et al., 2002b}

\begin{figure}
\includegraphics{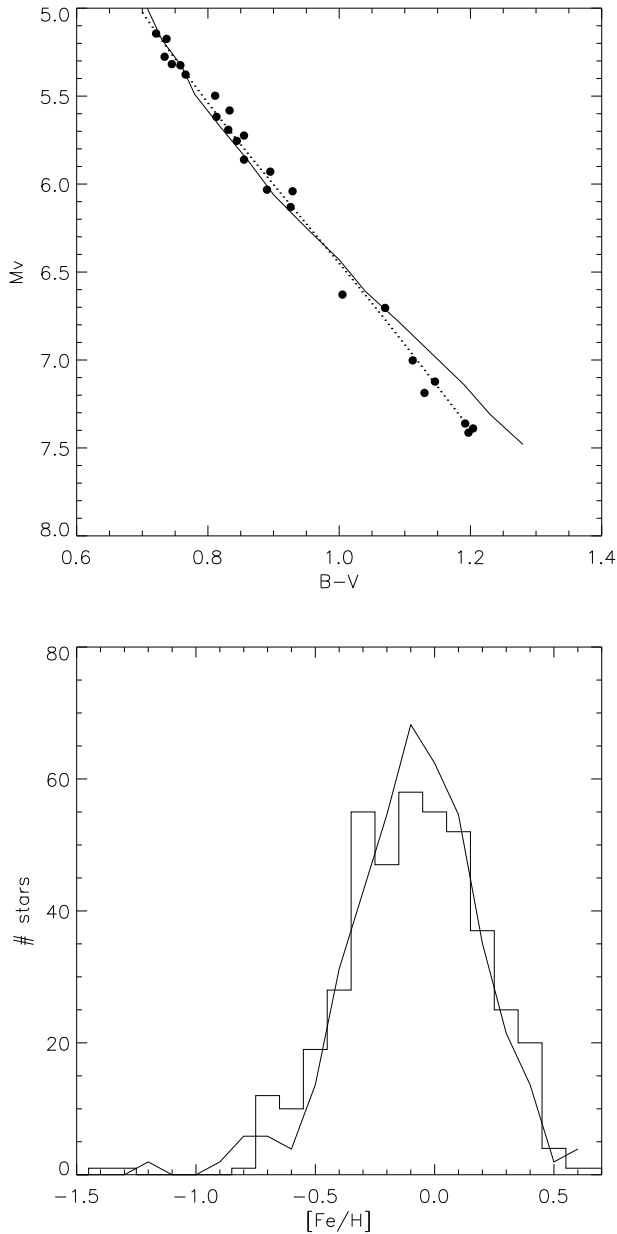}
\caption{ 
(a) The Hyades sequence from de Bruijne, Hoogerwerf, \& de 
Zeeuw (2001) (black dots), with the solar metallicity 
isochrone from Kotoneva et al.  (2002a).The dotted sequence is a fit to the 
observed Hyades sequence. At 0.75$<$B-V$<$1.07, the isochrone is very close
to the Hyades fit, and below it outside these limits.
(b) The histogram  of the stars selected by Kotoneva  et al. (2002b), with the
metallicity calculated using the Hyades
sequence  as the  reference (limited to  B-V$<$1.3). It  agrees well  with the
distribution given in Haywood (2002, Fig. 6), solid curve.  
}
\label{koto}
\end{figure}

There is a difference of 
about 0.2 dex between the distribution \cite{02KOT_EA2} and that of Haywood (2001).
We now try to clarify this point.
\cite{02KOT_EA2} assert that the calibration used in Haywood (2001)
is biased, based on a (B-V,[Fe/H]) plot (their Fig. 8) taken 
from our sample of long-lived dwarfs (Haywood 2001).
However, judging a calibration from a set of selected long-lived dwarfs is strongly 
misleading,  because  it shows  only  stars that are chosen on the basis 
of their colour being redder than a given isochrone.
This process imposes a limit that, varying with metallicities, is colour-dependent. 
This limit is visible on their Fig 8, where they show our sample of long-lived stars.
It is only an effect of the selection of long-lived dwarfs, not of the metallicity calibration.
This is the sense of the 'bias' seemingly detected by  Kotoneva et al. (2002b), and which 
therefore has nothing to do with a bias in the metallicity calibration we used.

Although their explanation is incorrect, \cite{02KOT_EA2} conclusively find
a discrepancy of about 0.2~dex between their calibration and the one we adopted, 
evaluated with 104 common stars.
What can be the origin of this  discrepancy ?  We have already commented
on our metallicity scale in Haywood (2002) and nothing new in Kotoneva et al. (2002b)
indicates a possible problem of our calibration (see also a discussion of our results 
by Taylor \& Croxall 2005).
Could the problem originate in the metallicity calibration of Kotoneva et al. ?
Figure~\ref{koto} shows
the solar metallicity isochrone of  \cite{02KOT_EA2} they
take as a reference scale, which difference of  absolute magnitude
to a given   star   is used as a metallicity indicator.
On the same  plot, we  also show  the Hyades  sequence as
given by de Bruijne et al.  (2001), together  with a  fit representing
this sequence. Since the Hyades cluster has a metallicity of +0.14~dex,
we would expect its sequence to be systematically 0.12 magnitude above
the isochrone (according to the  relation between metallicity and  the
difference  in magnitude of Kotoneva  et al. (2002a)).   As can be seen
from the plot, this is not the  case : the Hyades sequence is systematically
{\it below}   the isochrone  at $\rm  B-V>0.95$    and $\rm
B-V<0.75$, and the difference is much less than 0.12 magnitude between
these  limits. That   means their isochrone is probably in error, and it
follows that the calibration of Kotoneva et al. will
systematically underestimate the  metallicity  of the stars  over  the
whole  colour range  (by  about  0.15 dex), and most  severely  at $\rm
B-V>1.0$. 

Although their solar metallicity  isochrone is incorrect when compared
to the Hyades  sequence, it seems to fit well the  HR diagram of field, 'calibrating' 
stars, selected on the basis of photometrically determined
metallicity (Fig.  9 of  Kotoneva et al.  (2002a)).  How can  this be
understood ?   Kotoneva et  al.  do not  fit their isochrone  to solar
metallicity    stars {\it per se},    but     to    stars    selected    to    have
$0.1<[Fe/H]_{photo}<0.3$.  The argument  assumed by  the authors
being  that,  due to  observational  errors  in  the determination  of
photometric   metallicities,   and    the   peaked   distribution   of
metallicities,  stars  in  the  metal-rich  part  of  the  metallicity
distribution  (at   solar  metallicities,   in  this  case)   will  be
contaminated   by   the   lower,   dominant,   metallicities   (at
[Fe/H]$\approx$-0.2 dex, in their study).  To correct  for this effect,
and  to select stars  which have  a metallicity  which is  truly 
solar, the  authors select stars showing photometric metallicities
in the  range $0.1<[Fe/H]_{photo}<0.3$.  This  process meets two
difficulties  : First,  the correction  can be  safely  evaluated {\it
only}  if the  metallicity distribution  is  known a priori,  which is  a
dubious  method since the  metallicity distribution  is what they
want  to determine.   Second, if  the solar  neighbourhood metallicity
distribution peaks  at [Fe/H]$\approx$0.0, and not -0.2  dex (and assuming a symetrical
distribution), then the
correction  to  apply  when   selecting  solar  metallicity  stars  is
zero. If  this  is the  case  (Haywood 2001),  it
implies that the 'calibrating' stars in \cite{02KOT_EA2} are, in the
mean, genuine  high metallicity stars.  Unfortunately,
very few of  the 26 calibrating stars of \cite{02KOT_EA2} 
have a spectroscopic metallicity  in the catalogue  of \cite{01CAY_EA}.
However, HIP  99825 has a  metallicity -0.09 dex (+0.05,  Israelian et
al. 2004);  HIP 58345 has  a metallicity of  +0.16 dex; HIP  19788 has
+0.04 dex; HIP 15919 +0.26 dex and +0.33 dex, HIP 74135, 0.16 dex.

It is also possible to compare the metallicities in the whole catalogue
of Kotoneva et al. (2002b) (not only the calibrating stars) with spectroscopic measurements.
For the 31 stars in common with the catalogue of \cite{01CAY_EA},
the mean difference is 0.107~dex, in the sense that \cite{02KOT_EA2} 
underestimate the metallicities.
Using the newly published data  by Kovtyukh, Soubiran \& Belik (2004), 
we found that, for 34 stars  in common
between these two studies, the mean difference 
is also at 0.11 dex. Laws et al. (2003) find a similar difference between their
spectroscopic values and the metallicities of \cite{02KOT_EA2}. 
The great majority of these common stars have B-V$<$1.1.
The coincidence between their isochrone and the Hyades
sequence in this same colour interval (see Fig. \ref{koto}) 
implies a similar mismatch in metallicities. At B-V$>$1.1, 
the difference rises to larger values (implying underestimates
probably larger than 0.2 dex), their reference 
isochrone being above the Hyades sequence.
As a last check, we have computed the metallicity of the Hyades
stars using the absolute magnitudes derived by de Bruijne et al. (2001)
and the calibration of Kotoneva et al. (2000b, eq. 4). For the 23 members 
with 0.8$<$B-V$<$1.2, the derived mean metallicity is +0.045 dex, while restricting
the selection to 0.9$<$B-V$<$1.2  (13 members) yields -0.03 dex.
This amounts to 0.095 dex and 0.17 dex differences if the spectroscopic 
metallicity of the Hyades is 0.14 dex, and is consistent with our comments of the
metallicities of field stars.

The method adopted by Kotoneva et al. is however a useful alternative to 
classical metallicity indicators in a colour range where metallicities
estimates are notoriously difficult. 
An interesting exercise 
is to recalculate the metallicities in the sample of Kotoneva et al. (2002b),
using the  Hyades as the reference sequence.
This has been done using $\Delta$M$_v$ calculated for each star with 
the polynomial fit as the Hyades reference for B-V$<$1.3 (the limit of our polynomial). 
The calibration of Kotoneva et al. (2002a) is now modified to :\\

[Fe/H]=1.185$\Delta$M$_v$ +0.14\\

The new metallicity histogram of the star sample of Kotoneva et al. is 
plotted on Fig. \ref{koto}(b), and shows that the distribution
is now fully compatible with our own result (\cite{01HAY}).

Finally,  Kotoneva et  al. (2002b) cite  the  study of  \cite{97ROC_EA1}  as confirming  their results,  in contradiction with  our work.
The study of \cite{97ROC_EA1} is based on the calibration
of Schuster \& Nissen (1989),  which, for K  dwarfs, is known  to produce a strong  bias on
solar  and   super-solar  metallicity  stars,  when   applied  on  the
Str\"omgren photometry  of Olsen. (see Twarog, Anthony-Twarog, \& 
Tanner 2002; Haywood 2002).\\

\begin{figure}
\includegraphics{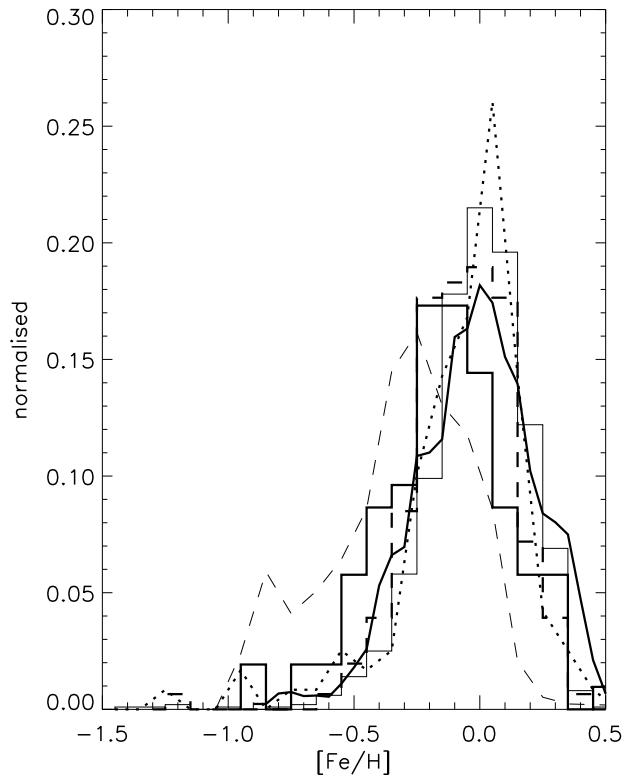}
\caption{Metallicity distributions for the samples discussed in section 2.
Haywood (2001) : thick curve;
Kotoneva et al. (2002), dashed curve;
Luck \& Heiter : dotted curve; 
Valenti \& Fischer (2005), whole sample : thin line histogram; 
Valenti \& Fischer (2005), 18pc sample :dashed histogram;
Allende Prieto et al. (2004) : thick line histogram.
}
\label{gdwarfcomp}
\end{figure}
\subsubsection{Spectroscopic samples}

Three different spectroscopic surveys of solar neighbourhood stars 
have started in the recent years by  Allende Prieto et al. (2004),
Luck \& Heiter (2005) and \cite{05VAL_EA}.
There is about 0.10 dex offset between our metallicity distribution
and that given in the 15pc sample of Allende Prieto et al. (2004), in
the sense that we give higher metallicities. On the contrary, the work
of Luck \& Heiter (2005) and \cite{05VAL_EA} agree well with our 
own study, with peak in the distributions around [Fe/H]=0.0~dex.
This is perhaps not surprising, since  Allende-Prieto et al. (2004) use 
Str\"omgren photometry and the calibration of Alonso et al. (1996) to derive effective
temperatures, which scale is known to be offseted by 100-150K compared to spectroscopic 
scales (see section 3.2 below).
To summarize, we plot on Fig. \ref{gdwarfcomp}
the distributions of these different studies, showing that there is good overall
consistency between the spectroscopic surveys (keeping in mind our comment 
about the Allende Prieto et al. sample) and Haywood (2001). 
(Note that the sample of Haywood (2001)  on  Fig. \ref{gdwarfcomp} 
is limited to long-lived dwarfs. However, comparison of the long-lived dwarfs metallicity distribution  
and the original sample in Haywood (2001) shows that the 2 distributions peak at the same metallicity.)
The shift of about 0.2 dex noted for the \cite{02KOT_EA2} sample is well in evidence.

\subsection{Scale height corrections}

\begin{figure}
\includegraphics[width=9cm]{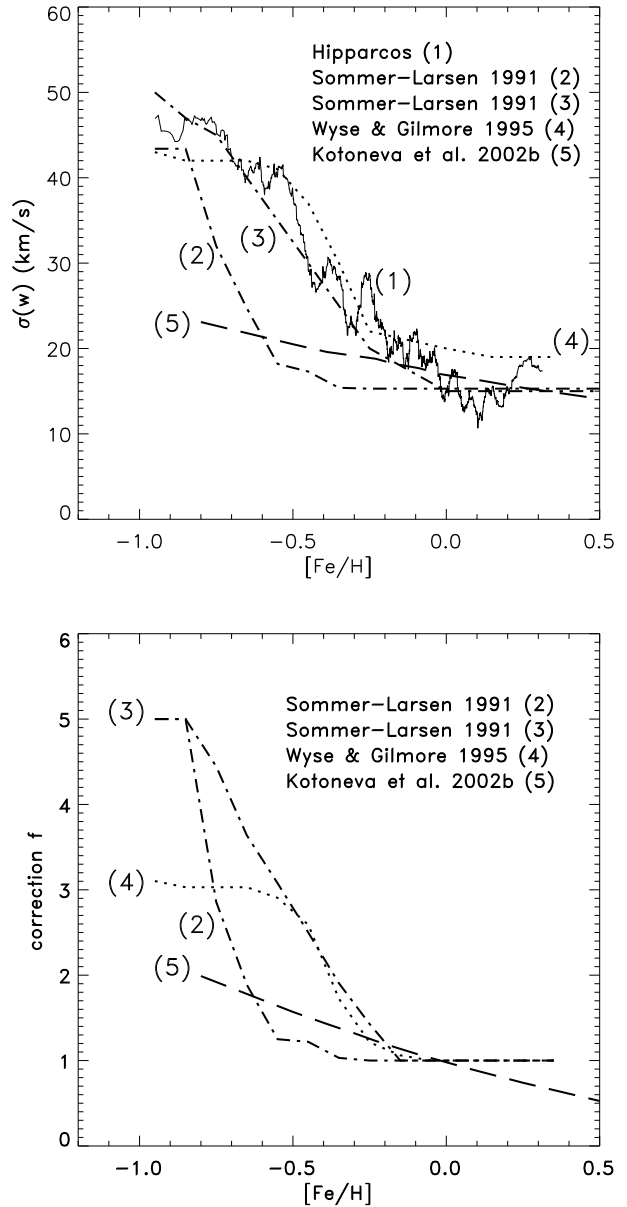}
\caption{(a) [Fe/H]-$\sigma_W$ relations used in different
studies to correct the local metallicity distribution. The black curve is 
the relation obtained over 3140 stars with metallicity from Geneva photometry
and W from Nordstr\"om et al. (2004). 
(b) The corrections for these same studies,  normalized at the solar metallicity. 
}
\label{scalecor}
\end{figure}

Because old stars make a kinematically hotter component than 
the young disc, they display a broader vertical density distribution
in a given potential. Their Galactic plane density is correlatively
lower, and they are under-represented in local samples. This differential
(age) effect biases  all estimates of the density and must be 
taken into account.
The corrections are usually calculated using a model of the vertical
structure of the Galactic disc related to the metallicity distribution
through a metallicity-vertical velocity dispersion ($\sigma_W$) relation. 
Figure \ref{scalecor}(a) shows various
[Fe/H]-$\sigma_W$ used in different studies for correcting the 
observed distributions. 
As can be seen, there are wide variations in these scale heights, the important
point being the amplitude between the minimum and maximum of $\sigma_W$.
Plot (b) shows the corresponding corrections adopted by these studies.
The same corresponding large spread is seen, from about 
2 for \cite{02KOT_EA2}, and 5 for the option (line 3 in Fig.3b) proposed by Sommer-Larsen (1991).
The result of \cite{02KOT_EA2} is somewhat surprising, because 
the thick disc is expected to contribute (if not to dominate) at [Fe/H] between [-0.6,-0.8] dex, 
while their $\sigma_W$ value seems to indicate a pure disc sample.

On Fig \ref{scalecor}(a), we show (labelled Hipparcos (1) on the plot) 
the running dispersion of a
sample of 3140 stars with Geneva photometry and 
well defined kinematics from the sample of Nordstr\"om et al. (2004).
Although the sample is not complete in distance, it is sufficiently 
large that undersampling uncertainties are small, the metal-weak 
([Fe/H]$<$-0.5) part of the sample representing a fair 9\% of the sample.
The vertical velocity dispersion decreases from about 46~km.s$^{-1}$ to 
about 13~km.s$^{-1}$ at [Fe/H]=0.1 dex. The rise at supersolar  metallicities
is due to the inclusion of old, metal rich objects. 
A similar result is obtained with the sample of 
Soubiran \& Girard  (2005). Selecting from their sample stars they
identify as thin and thick disc objects (excluding  members
of the Hercule stream), the dispersion varies from about 42km.s$^{-1}$
at [Fe/H]$<$-0.65 dex to about 13 km.s$^{-1}$ between -0.1$<$[Fe/H]$<$+0.1~dex.
These new data clearly illustrate that the largest (NR option in his paper) 
scale height corrections  of Sommer-Larsen (1991) are very near to the observed 
solar neighbourhood metallicity-$\sigma_W$ relation.

\begin{figure}
\includegraphics[width=9cm]{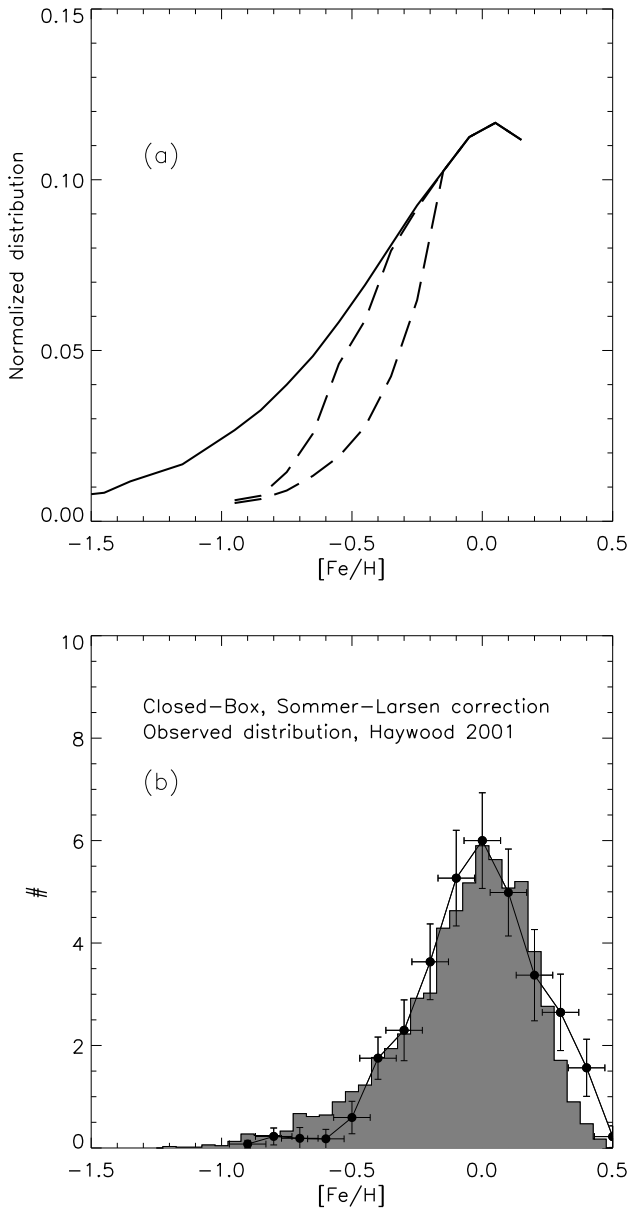}
\caption{The closed-box model metallicity distribution (continuous curve)
  and the (local) volume density closed-box model, corrected using the scale height corrections
of Sommer-Larsen (2) and (3) of Fig. \ref{scalecor}.
}
\label{closedbox}
\end{figure}

\subsection{Comparison with the closed-box model}

Having determined the scale heights, we want to compare the observed and
model metallicity distributions. Usually, scale height corrections are
applied to the observed metallicity distribution. However, due to the small
number statistics of our original sample (Haywood 2001) at [Fe/H]$<$-0.5 
and the uncertainties in the metallicity scale, this procedure will systematically 
amplify poisson variations in 
the observed metallicity distribution. We prefer to bring the model
nearer to the observations and apply the scale height corrections to
the predictions of the closed-box model. 
Fig. \ref{closedbox} shows the closed-box  model prediction, corrected to give 
volume density distribution and convolved with 0.1 dex gaussian errors, 
 together with  the long-lived dwarf metallicity distribution of Haywood (2001).
As is clear from the figure, there is no marked disagreement between 
the two distributions. There is a slight overestimate of the model at
[Fe/H]$<$-0.5, but this is hardly significant, given the uncertainties
in the detailed profiles of the scale height corrections and metallicity 
scales.
It is clear from what has been said that the match between the closed-box
model and the observed distribution is satisfactory only because the thick 
disc has been included in the comparison.

\subsection{Discussion}

Formally, we have shown that, using the same procedures as those found in studies of the 'G-dwarf problem' 
and strictly restricting our case to the  fit of the metallicity distribution, 
the most up-to-date data don't lead to a 'G-dwarf problem'. Before the significance
of this 'no-G-dwarf problem' can be discussed (section 5), it is important to point out several inconsistencies that 
have hampered a clear assessment of this question in the last fifteen years or so.

Our result, based on local data, is compatible with arguments of Galactic structure evocated
in the introduction that, given the properties of the thick disc (scale height and local density), this population should 
contribute (at  -1.0 $<$[Fe/H]$<$-0.5 dex) a rough 15-20\% of the - corrected for volume effect - 
metallicity distribution.
This point is barely discussed in the literature on the subject, and we may ask 
how the thick disc contribution is dealt with  in the numerous studies published on the subject.
In papers deriving the observed metallicity distributions, thick disc stars
are not disregarded and therefore should have significant contribution when volume corrections
are considered. However, 
the characteristics of the derived metallicity distributions do not meet the expectations just mentioned.
In the distribution by J{\o}rgensen (2000) (the characteristics of which are the nearest to our 
own distribution), the contribution of the stars with -1.0 $<$[Fe/H]$<$-0.5 dex is left almost completely 
unmodified by the scale height corrections, representing from about 5 (uncorrected) to 7\% (corrected) of the total distribution.
In  Kotoneva et al. (2002b),  and assuming that their metallicity scale has to be moved up by
+0.2 dex (see section 2.1.1), the contribution of stars in this same metallicity interval, 
after correction for  volume effect,  is only 12\%. 

In papers where models are designed  to fit the observed distribution, the thick disc is usually
acknowledged as a genuine Galactic population. 
In most cases however, models would not be able 
to fit both 15-20\% of the distribution between  -1.0 $<$[Fe/H]$<$-0.5 dex  and a peak centred at 
[Fe/H]=0.0 dex. Limiting our census to the most recent models, in Chiappini, Matteucci \& Gratton (1997), the percentage of stars within -1.0 $<$[Fe/H]$<$-0.5 dex 
given by the model is about 12\%, and the peak is centred on [Fe/H]=-0.1 dex.  Shifting it to solar
metallicity would make the 'thick disc' even less conspicuous. Alib{\'e}s, Labay, \& 
Canal (2001)  metallicity distribution is similar.
In Renda et al. (2005), the material produced by the model within this metallicity range
matches the lower thick disc estimates (16\%) however, the peak metallicity is set at [Fe/H]=-0.2 dex, 
to comply with the observed distribution of Kotoneva et al. (2002b). 

In most (if not all) recent studies of the dwarf metallicity distribution the thick disc 
contribution seems to have been overlooked both by models and in papers derivating the metallicity
distribution from observation.

\section{The age-metallicity relation}

The age-metallicity distribution is considered a loose constraint of 
the Galactic chemical evolution, because of the lack of clear trends 
and the discrepant results that have been obtained.
Suggestions for the existence of such relation have been given by \cite{72POW} and Hearnshaw (1972),  
followed by \cite{80TWA2}, while Carlberg et al. (1985) found a rather flat and dispersed relation.
More recent work (Edvardsson et al. 1993) have also cast some doubt on the existence of a clear 
correlation between age and metallicity. On the contrary, a study based on chromospheric 
activity  by \cite{00ROC_EA1} found a rather tight correlation. At the same time, 
Garnett \& Kobulnicky (2000) demonstrated that the large scatter in metallicity in Edvardsson et al. (1993)
is more likely to reflect the sample selection than real scatter in the local stars.
The overall impression is therefore that the issue has not been settled with pre-Hipparcos data.
There are good reasons to  this failure that were already known to Tinsley (1974), 'that 
stars of all ages have a considerable dispersion in Z', while 'the mean value is a very slow increasing
function of birth epoch'.
The publication of the Hipparcos catalogue (1997) and the Geneva-Copenhagen Survey (hereafter GCS) 
(Nordstr\"om et al. 2004)  have been important milestones for the study of the solar neighbourhood, 
so that we may ask: did we get closer ? 
Feltzing et al. (2001) and \cite{04NOR_EA})
have provided the most comprehensive studies, with samples of about 5000 stars or larger
and detailed age determinations, pointing to a similarly dispersed age-metallicity distribution.
\cite{04PON} focused on a restricted but careful age determination and discussed a 
possible correlation in the Edvardsson et al. (1993) sample.
\cite{04PON} have demonstrated that classical isochrone dating
method is subject to biases that naturally arise in regions of the HR diagram where
the effects of ages on the atmospheric parameters become small, even when 
the calibration scales  providing these atmospheric parameters are correctly set.

We take here a different view, and try to evaluate how ages are affected when 
such calibrations are biased.
We show how explicit biasing of the atmospheric parameters
can lead to structures and spurious patterns in the AMD, even at ages below 3 Gyrs.
Then we focus on analysing the recent AMD of Feltzing et al. (2001) and Nordstr\"om et al. (2004).
Although we use mostly the GCS catalogue from Nordstr\"om et al. (2004), 
our conclusions are applicable to Feltzing et al. (2001), since Nordstr\"om et al. (2004) have used essentially
the same input data and calibrations, and obtained a similar AMD.

\subsection{Modeling biases in the age-metallicity distribution}

\begin{figure*}
\includegraphics[width=18cm]{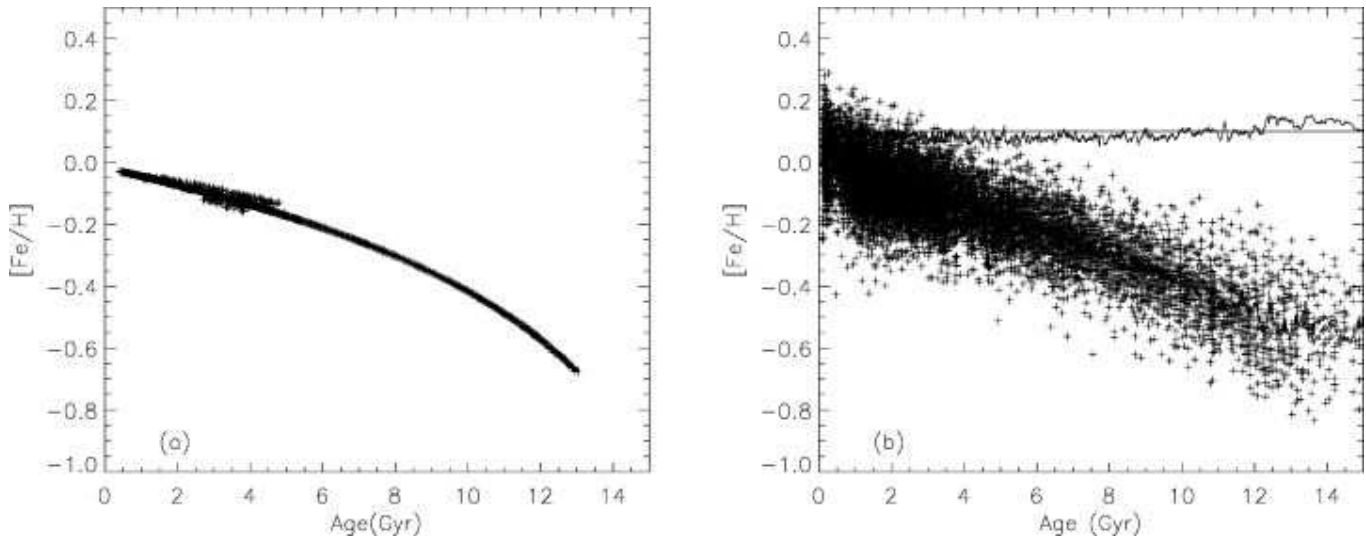}
\caption{
(a) The adopted age-metallicity relation (AMR) as it is recovered when no bias on metallicity or effective temperature
are introduced. The only significant deviation from the original relation 
happens for stars younger than 5 Gyrs, due to the overlap of isochrones in the hook region.
(b) The AMR relation recovered assuming 0.1 dex dispersion on metallicity and 50K on effective
temperature. No significant dispersion is added by recovered ages.}
\label{agemetsimu}
\end{figure*}

\begin{figure*}
\includegraphics{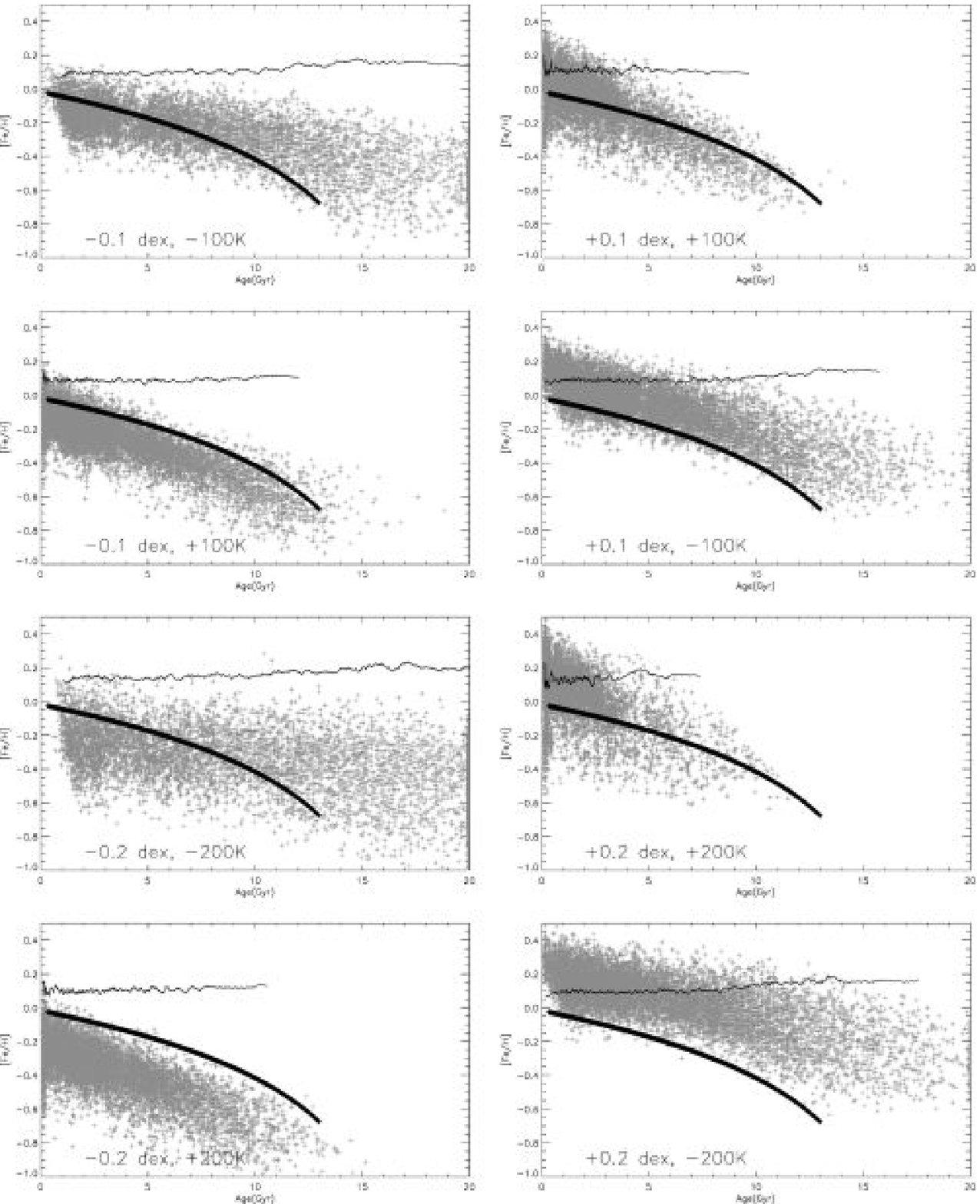}
\caption{Theoretical age-metallicity relation (thick curve), and samples of about 10000 'stars' 
whose ages have been recovered assuming the various systematic biases given on each plot.
We also assume gaussian random errors on metallicities of 0.1 dex, and 50K on temperatures.
Thin curve on each plot is the metallicity dispersion calculated over 100 points.}
\label{allbias}
\end{figure*}

In order to evaluate the effect of systematic and random errors on the 
determination of age in photometric samples, we apply the following simple procedure.
We assume a model AMD, from which we generate a sample
of about 10000 points in the HR diagram, using a set of isochrones (\cite{03YI_EA}). 
The sample of simulated 'stars' is limited to absolute magnitude $2<M_v<5.5$, and 0.3$<$B-V$<$1.0.

Systematic biases and random errors are then introduced on 
the atmospheric parameters $\rm T_{eff}$ and [Fe/H], from which we 
try to recover the age using the isochrone fitting method (and the same 
set of isochrones).
We can then compare the resulting ages with the initially assumed relation.
Although we test what can be expected from the data only, with no interference from the stellar models,
it must be kept in mind that they are a source of uncertainty that is difficult to evaluate.
Stellar models certainly can generate similar systematic effects as those studied here, by 
being too hot, too cold, etc. Also, the test is a simplified one, 
and does not take into account effects such as unresolved binaries, $\alpha$ enrichment
of the stellar models, error on absolute magnitude, etc.

The age search is made by scanning a set  of isochrones at the metallicity of the star to minimise 
the $\chi^2$  quantity :\\

\begin{equation}
 \chi^2 =  (M_v^o - M_v^m)^2 +16*(LogT_{eff}^o-LogT_{eff}^m)^2
\end{equation}

which is the formula introduced by Ng \& Bertelli (1998) and 
also used by Feltzing et al. (2001).
The subscript $o$ is meant for observations and $m$ for model.
The model and recovered age-metallicity relations are shown on Fig. 5 assuming 
different biases on atmospheric parameters, as described now.

 (a) We first assume a theoretical AMD, represented by a single curve, with no intrinsic dispersion
in metallicity at a given age, no bias on atmospheric parameters. 
The aim here is to evaluate the uncertainties due to the method. 
The result is given in Fig. \ref{agemetsimu}(a), and shows that besides a few points at ages smaller than 5 Gyrs,
due to overlapping isochrones in the hook region, the method is satisfactory.
This feature disapears when the hook itseft disapears, at ages 
greater than 5 Gyrs. 

(b)  The ages are now recovered assuming 0.1 dex 
and 50 K dispersion in the input metallicities and effective temperatures in order
to simulate the effect of a reasonably small observational scatter.
The result is given on Fig. \ref{agemetsimu}(b). The curve gives the dispersion 
in metallicity as a function of age, and shows that there is no increase of the 
dispersion in metallicity at ages less than 12 Gyrs. 
Only 5\% of the stars have a new age differing by more than 0.1 Gyr from their true age, 
hence, if no other source of error are present, the determination of the AMD can be made with
confidence.

(c) The AMD have now been biased systematically in temperature and metallicity, while 
the random errors are the same as previous values (0.1 dex in [Fe/H], 50~K in T$_{eff}$). 
The results are shown on Fig. \ref{allbias} for various combinations of biases. 
It can be seen that combining moderate random errors and biases easily produces
significant deviations from the original relation, with ages modified by several Gyrs (up to almost
10 Gyrs in case of underestimated effective temperatures). Although biases in metallicity 
affect the shape of the AMD, biases in effective 
temperature have the most dramatic effects, easily creating young metal-poor 
or old metal-rich stars with just a hundred degrees bias on effective temperatures.
Another consequence is that, starting from an AMD with no intrinsic scatter, 
a dispersion in metallicity of about 0.15-0.20 dex is easily reached with a 
combination of (limited) observational random errors and systematic bias on 
effective temperatures, in particular for old stars.
It implies that a combination of even relatively minor systematics on effective temperatures
and metallicities can seriously affect the general morphology of the AMD.

\subsection{Uncertainties in atmospheric parameters}

In order to assess how robust is the determination of the AMD from large Str\"omgren
photometric surveys, we need to examine the determination of atmospheric parameters from
photometric indices.
In our discussion, we focus on effective temperatures and metallicities, because
the errors on these two parameters dominate the final error on age determination.

\subsubsection{Effective temperatures}

\begin{figure} 
\includegraphics[width=8cm]{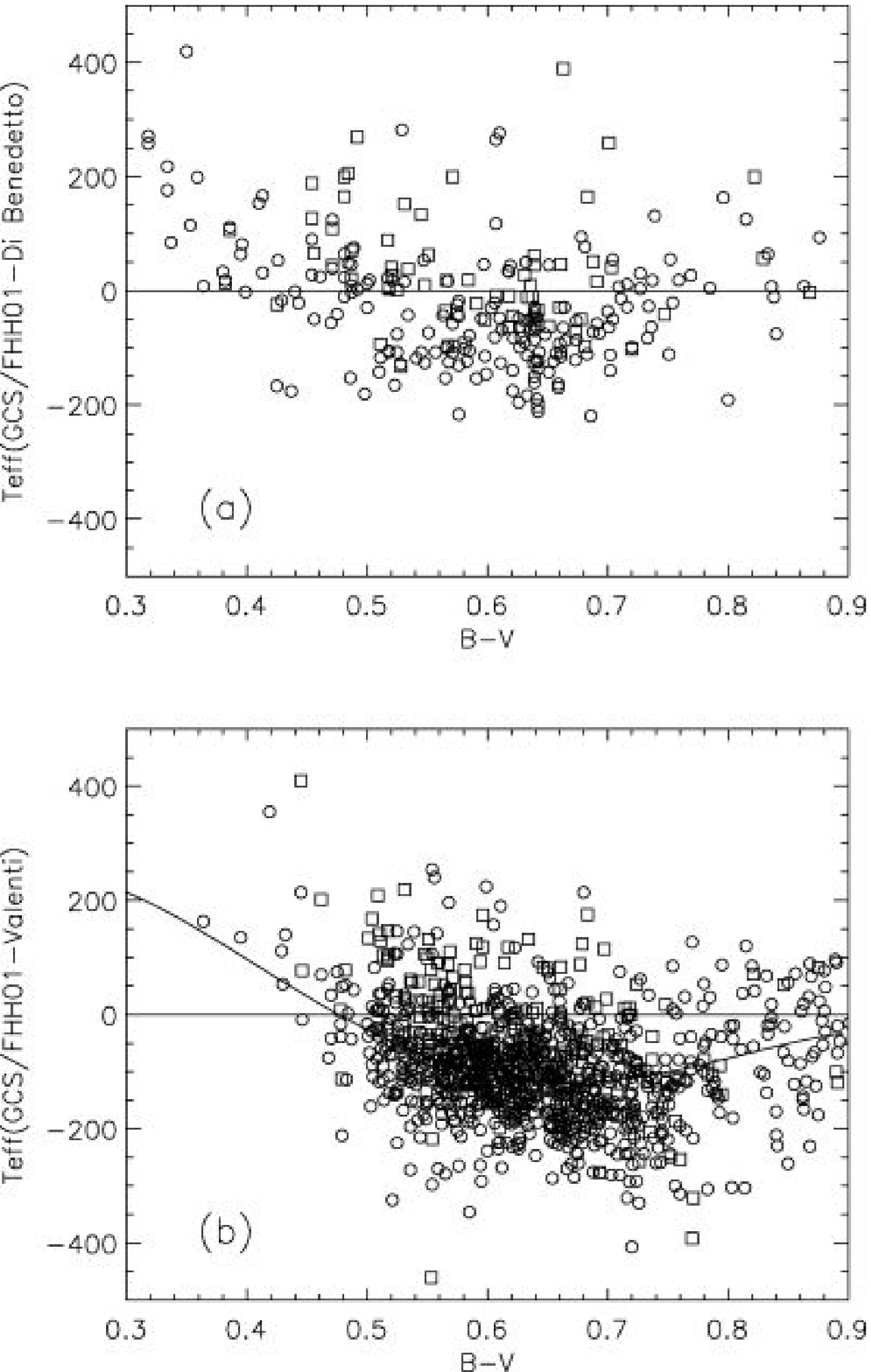}
\caption{(a) Differences between effective temperatures from the Geneva-Copenhagen Survey and 
the catalogue of Di Benedetto (1998) (circles), for 179 stars with 0.3$<$B-V$<$0.9 and Feltzing et al. (2001) for 
70 common objects (squares).
(b) 740 stars
within the same limits from Valenti \& Fischer (2005). The curve is a fit to the combination
of the 2 datasets. It is used for correcting the effective temperatures in the GCS catalogue. 
}
\label{diBe}
\end{figure}

The two studies of Feltzing et al. (2001) and Nordstr\"om et al. (2004) use the scale of 
\cite{96ALO_EA1} that gives effective temperature as a function of the b-y Str\"omgren index 
and metallicity. 
On the temperature range 
between 5250 and 6250 K, where most datable stars are found, and for $\approx$700 stars in
common, the temperatures of Nordstr\"om et al. are 136 K lower than the spectroscopic temperatures of Valenti \& Fischer
(2005). Valenti \& Fischer have 94 K difference with Allende Prieto et al.
(2004), who also based their temperature scale on that of  \cite{96ALO_EA1}. Luck \& Heiter
note a  difference of 158 K  between their spectroscopic scale and that of Allende Prieto et al. (2004).
Santos et al. (2004), have noted also an offset of 139K in the same sense between their 
spectroscopic determinations and the temperature scale 
of \cite{96ALO_EA1}, and similar result are also found in Takeda et al. (2005).
Ram\'irez \& Melendez (2005b) revised the scale of Alonso et al. (1996) to produce 
essentially the same scale.
There seems to be a real dichotomy of the order of 100-150K between the photometric and spectroscopic
effective temperature scales, and we take the view that this is the amount of possible 
systematic errors on effective temperatures. In order to know which scale is correct, 
we need an independent estimator of the effective temperatures. 
This can be provided by effective temperatures derived from the Stefan-Boltzmann relation, 
measurements of stellar angular diameters and bolometric fluxes, or a scale based 
on such basic data. This is provided for example by Di Benedetto (1998), which gives
a T$_{eff}$-V-K relation calibrated on effective temperatures  derived from angular diameters.
We have cross-correlated the effective temperatures from the catalogue of Di Benedetto (1998) 
with the GCS catalogue and found
190 common stars with B-V between 0.3 and 1.0. The differences in effective temperature 
between the two scales are plotted on Fig. \ref{diBe}(a), and confirm that, in the 
crucial range B-V=0.5-0.7, Nordstr\"om et al. (2004) underestimate effective temperature by
$\approx$100K. Fig. \ref{diBe}(b) shows the difference between the spectroscopic temperatures of
Valenti \& Fischer (2005) and the GCS catalogue for 740 objects. The discrepancies are very
similar, in extent and amplitude, to that of Di Benedetto (1998). We have estimated a mean correction
to apply to the GCS effective temperatures by fitting a polynomial to the joint datasets of Di Benedetto (1998)
and Valenti \& Fischer (2005). This fit is shown on Fig. \ref{diBe}(b).

\subsubsection{Metallicities}

\begin{figure}
\includegraphics[width=8cm]{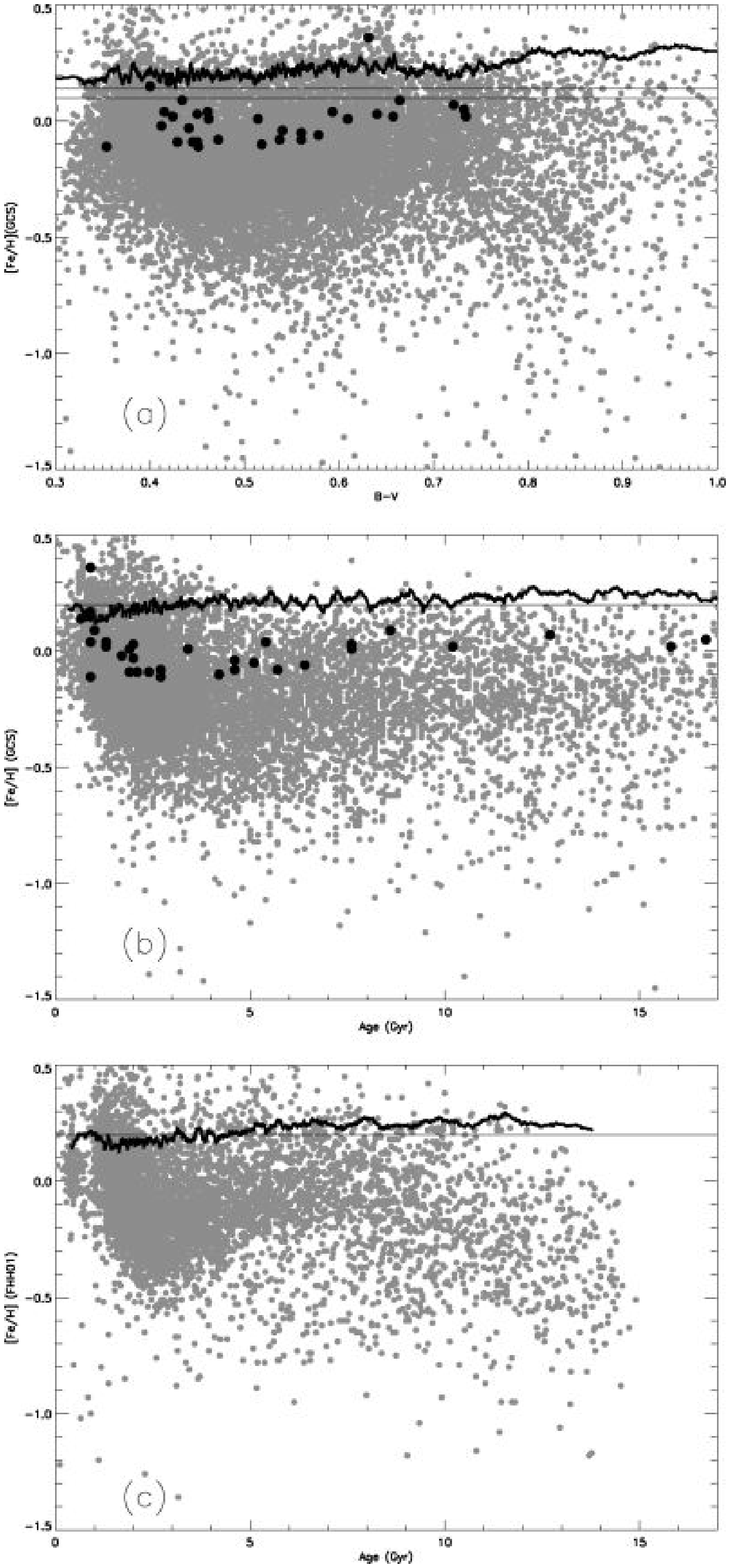}
\caption{{\bf (a)} (B-V,[Fe/H]) data from the catalogue of Nordstr\"om et al. (2004).
The 2 horizontal lines bracket the measured spectroscopic metallicity of the Hyades cluster (0.1-0.14 dex), 
while the dots are the Hyades stars from the catalogue of Nordstr\"om et al.
The thick curve is the metallicity dispersion calculated over 100 points.
{\bf (b)} The age-metallicity distribution for field stars and the Hyades from the GCS catalogue.
The wavy pattern seen in (a) is also reflected in the AMD (b). 
Inspection of the ages of individual Hyades stars in the catalogue shows that they 
correlate with B-V.
The thick curve is the metallicity dispersion calculated over 100 points. The thin horizontal
line is a guide to evaluate the dispersion.
{\bf (c)} The  age-metallicity distribution for field stars from Feltzing et al. (2001).
Patterns in the AMD (b\&c) are a combination of a defective metallicity calibration, the 'terminal age bias'
and systematics in the temperature scale.
}
\label{fehage}
\end{figure}

Discrepancies in the metallicity of solar neighbourhood dwarfs (Section 2) illustrate 
the difficulties of adopting a correct photometric metallicity scale. 
Establishing a calibration of photometric metallicities meets several difficulties, 
one of which are the systematic differences between various spectroscopic data sets.
Although there are several possible causes for mismatch between spectroscopic scales, 
it is known that effective temperature differences of about 100 K induce an offset of about
0.06-0.07 dex. This is an obvious possible explanation for the differences between the datasets
of \cite{05VAL_EA} and Allende Prieto et al. (2004).
A second problem are biases that depend on
colour, which are usually not seen by visualising simple correlations between
spectroscopic metallicities and photometric estimates.
It is crucial that such biases be eliminated, because they would easily affects the shape of the 
AMD, since colour on the main sequence is strongly correlated with age. 

It has been shown that such biases operate in the Str\"omgren calibration of metallicity
from \cite{89SCH_EA1} for G {\it and} F dwarfs (Haywood 2002), although 
the bias may not be detected in simple spectroscopic-photometric metallicity correlations.
This bias is difficult to correct, and it is not eliminated from the 
sample of Norsdtr\"om et al. (2004), although the authors claim to have proceeded to new adjustments.
A general underestimate of the metallicity is apparent in \cite{04NOR_EA} when 
their catalogue is compared with spectroscopic metallicities. For example there is a mean difference 
of -0.075 dex with \cite{05VAL_EA} (834 stars). A similar offset is observed 
with the compilation of \cite{01CAY_EA}, varying from -0.06 between 0.4$<$B-V$<$0.5, to
zero above this limit. These are general values, but a detailed comparison 
reveals more problematic differences.

Fig. \ref{fehage}a shows the (B-V,[Fe/H]) distribution from \cite{04NOR_EA}
for field stars, together with their metallicities for the Hyades stars.
It is known that the spectroscopic metallicity of the Hyades is usually
measured between 0.10 and 0.14 dex, as indicated by the 2 horizontal lines
on the plot. It is remarkable that, in the interval where most of the datable 
stars are found (B-V$<$0.6), the metallicity of the Hyades from Nordstr\"om et al. 
is underestimated by 0.15-0.2 dex. Although this problem could affect only the cluster members, 
the similar wavy behavior of the field star distribution and the Hyades 
leaves little doubts that the problem is general in the catalogue 
and indicates serious colour effects in the metallicity calibration.

\begin{figure}
\includegraphics[width=9cm]{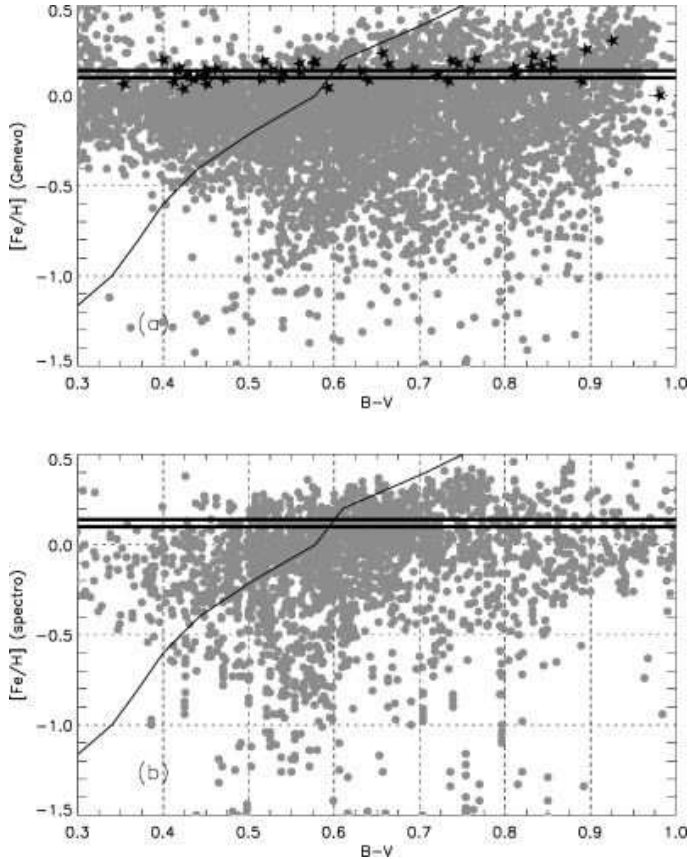}
\caption{(a) (B-V, [Fe/H]) distribution using Geneva photometric metallicities (about 5100 stars).
The thick horizontal lines bracket the spectroscopy metallicity of the Hyades (0.10-0.14 dex).
The star symbols show the metallicity of the Hyades according to the Geneva photometry.
The thin line gives the turn-off B-V colour as a function of metallicity for a 5 Gyrs isochrone. 
(b) (B-V, [Fe/H]) for spectroscopic metallicities from various sources.
The similarities between the 2 figures constrast with the differences with the data 
from Nordstr\"om et al. (2004) (Fig. 8a). Note in particular the metal rich stars at B-V$>$0.7, 
absent from Nordstr\"om et al. (2004). The wavy structure seen in Fig. 8a is not seen either.
}
\label{bvfehgen}
\end{figure}

This is confirmed by Fig \ref{bvfehgen}(a\&b), which shows the (B-V,[Fe/H]) distribution
obtained with two, independent but consistent, metallicity scales.
The first one is a set of stars with [Fe/H] derived from Geneva photometry (using metallicity calibration
from Haywood (2001) and Geneva photometric indices, it is designated herebelow as 'Geneva metallicities'), 
and the second is made of spectroscopic metallicities from various sources 
(Balachandran 1990; 
Edvardsson et al. 1993; 
Gratton, Carretta \& Castelli 1996;
Favata, Micela, \& Sciortino 1997; 
Feltzing \& Gustafsson 1998;
Chen et al. 2000; 
Fulbright 2000; 
Takeda et al. 2005; 
Bensby, Feltzing \& Lundstr{\" o}m 2003; 
Erspammer \& North 2003; 
Valenti \& Fischer 2005;
Woolf \& Wallerstein 2005
). 
These two distributions are very similar,  although the spectroscopic one is clearly
incomplete. Considering that the B-V scale is also an age sequence, with older stars 
being progressively included towards higher B-V, they present the following characteristics. 
The range of metallicities is clearly much narrower for blue, or young, objects.
This is due not only to the (expected) appearance of metal poor but also more
metal rich objects at redder colour. It is apparent that for the youngest (bluest) objects,
the upper metallicity is approximately that of the Hyades cluster (star symbols on Fig. 9a).
This implies that the progressive increase of the upper metallicity towards greater B-V in 
the sample is due  to the inclusion of progressively older stars, starting from the Hyades
metallicity at +0.15 dex and reaching +0.5 dex. This can be evaluated visually with the curve of 
turn-off points at different metallicities for a 5 Gyrs isochrone.
 We comment further on this point in section 4.
The second feature  of  importance is the fact that the youngest objects do show a significantly 
narrower range of metallicity than the redder, and, in the mean, older, objects. Note that
(1) this feature is robust since it is independent of any age determination (2)
it is contradicting the recurrent claim of a uniform dispersion at all ages that have been 
claimed in most recent studies of the AMD, and certainly points to a problem in the determination 
of ages. In section 4 of this paper, it is shown that new ages, based on revised atmospheric
parameters, are consistent with the (B-V,[Fe/H]) relation.

There are two distinct features that are easily seen when
comparing these  (B-V,[Fe/H]) datasets with \cite{04NOR_EA}.
The first one is the wavy behavior seen in Nordstr\"om et al. which is absent
in the other metallicity samples. The second is the near-absence of cold (B-V$>$0.65), 
super metal-rich ([Fe/H]$>$0.2 dex) stars in the GCS that are conspicious in Geneva photometric 
metallicity, and whose gradual appearance is clearly visible on the spectroscopic data (Fig \ref{bvfehgen}b)
(The small uprise of metallicities seen in Nordstr\"om et al. data at 0.60$<$B-V$<$0.65 
is due to biased metallicities, as will be shown later in this section).
This is mainly a sampling problem : 16\% of the sample have B-V$>$0.65 and [Fe/H]$>$0.2 dex in  Valenti \& Fischer (2005)
and the Geneva samples, 4\% in  Nordstr\"om et al.
Isochrones show that at [Fe/H]=0.3 dex and age $>$ 5 Gyrs, turn-off stars have B-V$>$0.67.
There are 17 stars in the GCS catalogue within these limits, and the same number in  
Valenti \& Fischer (2005), although it is more than 10 times smaller. 
Out of these 17 objects, only 2 have a GCS age greater than 5 Gyrs.
In view of these numbers, the scarcity of stars at [Fe/H]$>$0.2 dex and 4-5 Gyrs 
in the AMD from Nordstr\"om et al. (2004)
is more likely to be a selection effect than real depletion. 
This should be emphasized, since the authors insist that their
age-metallicity coverage is complete.

\subsection{The age distribution of Feltzing et al. (2001) and Nordstr\"om et al. (2004)}

Comparison between the AMD of Feltzing et al. (2001) and 
\cite{04NOR_EA} (Fig. \ref{fehage}) shows that more or less sophisticated dating
methods don't lead to drastically different results. Although the error analysis
is certainly different, the general characteristics of the AMD are the same in the two studies.
These are :
(1) a decrease of the mean metallicity with decreasing age between 10 and 5 Gyrs, 
(2) a clump of stars at age $<$ 3 Gyrs,
(3) the existence of young (age $<$ 5 Gyrs) metal-poor ([Fe/H]$<$-0.5 dex) stars and correlatively, the relative depletion of old metal-poor objects,
(4) a high dispersion at all ages.
We comment each of these features in turn.

\subsubsection{Decreasing metallicity with age}

The AMD of Nordstr\"om et al. (2004), as well as the one of Feltzing et al. (2001) shows a mean
decrease the metallicity between 8-10 Gyrs and 4-5 Gyrs, before a sharp rise at younger ages (Fig. 8bc). 
A similar decrease is seen in the (B-V, [Fe/H]) distribution of  Nordstr\"om et al. (2004) (Fig. 8a), between B-V=0.7 and 0.5.
This feature is reproduced identically on the Hyades stars (black dots),
so that it is most probably due to a defective metallity calibration.
The similarity between the 'wavy' pattern of the colour and age metallicity distributions
is obvious, and shows that the AMD reflects mainly metallicity
variations that are visible in the (B-V,[Fe/H]) plot.
Note that the interesting matter
here is the fact that the age-metallicity pattern of the Hyades reflects the field star sample. 
The large errors on individual ages of the Hyades stars are less surprising 
since these are near the zero-age main sequence, 
and therefore particularly sensitive to errors on the metallicities and effective temperatures.
The cluster stars confirm that the metallicity variations in age closely follow 
those seen as a function of B-V. The decrease in metallicity of the Hyades stars with age between 
9 to 5 Gyrs strongly suggest that the similar apparent decrease in field stars 
is only reflecting the same defect in the  metallicity calibration.

\subsubsection{The clump of stars at age $<$ 3 Gyrs}

\begin{figure}
\includegraphics[width=8cm]{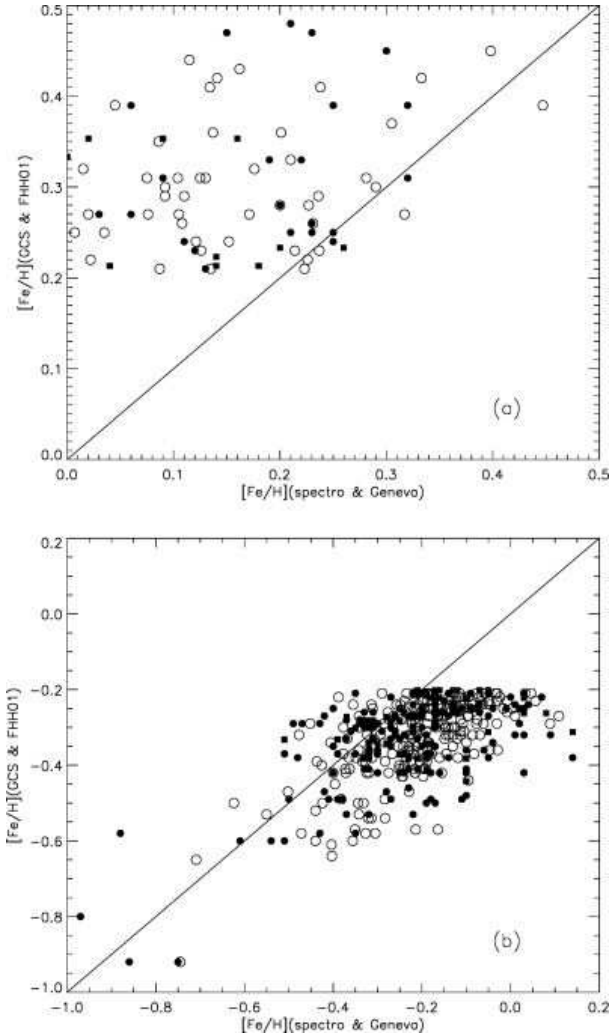}
\caption{(a) Spectroscopic data $vs$ stars that have [Fe/H] $>$ 0.2 dex and 
age $<$ 3 Gyrs in Nordstr\"om et al. (2004, GCS catalogue) (dots) and Feltzing et al. (2001) 
(filled squares).
When available, Geneva photometric metallicities (discussed in section 3.2.2) $vs$ 
Nordstr\"om et al. (2004) metallicities have also been plotted (open circles), 
showing the same trend. 
 (b) Stars that have [Fe/H] $<$ -0.2 dex and age $<$ 3 Gyrs in 
Nordstr\"om et al. (2004) and Feltzing et al. (2001). Symbols are the same as in (a).}
\label{biasedfeh}
\end{figure}

The most conspicuous pattern in the AMD of Nordstr\"om et al. and 
Feltzing et al., is the 'clump' of stars at  age $<$ 3 Gyrs, spanning more than 1 dex in metallicity, from
[Fe/H]$<$-0.5 to 
[Fe/H]$\approx$0.5. Is this feature real ? 
The sudden rise of the Hyades metallicity from -0.1 dex to 0.2 dex within 2 Gyrs, 
closely following that of field stars, suggest it is not. 
It suggests that the youngest field stars have overestimated metallicities, while those 
at age 2-3 Gyrs have underestimated metallicities.
We first focus on the young, metal rich objects. \\

(a) Overestimated metallicities\\

Nordstr\"om et al. (2004) suggest the super-metal rich young stars in the GCS catalogue 
could in part be giant stars misleadingly included
due to improper dereddening. However, half of the stars with [Fe/H]$>$0.25 dex and age $<$3 Gyrs 
in their catalogue have a parallax greater than 10 mas and are unlikely to suffer from 
important reddening problems. The other half is made of objects
whose absolute magnitude and colour show they are evolved main sequence or subgiant stars. 
While problems in reddening correction may be a valid explanation for these stars, it remains 
speculative. Comparison with spectroscopic metallicities suggests another explanation by showing that 
they have overestimated metallicities. 
Stars selected in the GCS  
with [Fe/H]$>$+0.2 dex and age $<$ 3 Gyrs  have been compared with spectroscopic 
values from the catalogue of \cite{01CAY_EA}, Feltzing \& Gustafsson (1998) and \cite{05VAL_EA}
on Fig. \ref{biasedfeh}. 
Although only 26 objects (8 in Feltzing et al. (2001)) have been found, this is sufficient to
confirm that photometric metallicities are overestimated. All 26 objects are either main sequence or
subgiant stars within 100pc, most at nearer than 70pc, so that reddening problems are excluded.
Note that all measurements available for a 
given star in  Cayrel et al. (2001) are plotted on these figures.
In addition, the GCS catalogue has been correlated, for the same selections, 
with the sample of Geneva photometric metallicities discussed in section 3.2.2. 
38 stars were found, and are plotted as open circles on Fig. 10, showing the same
trend as the spectroscopic data.\\

(b) Underestimated metallicities\\

At 2-3 Gyrs, the Hyades suggest that metallicities  in Nordstr\"om et al.  and Feltzing et al. are 
probably underestimated. Again, the confirmation 
comes from spectroscopic determinations, as can be checked in Fig \ref{biasedfeh}(b): stars selected in the catalogue of Nordstr\"om et al.
with [Fe/H]$<$-0.2 dex and age $<$ 3 Gyrs  are compared with spectroscopic 
values from the catalogue of \cite{01CAY_EA}. There are 78 objects in common between the two
datasets (24 for the dataset of Feltzing et al. (2001)). 
Figure \ref{biasedfeh} shows all values available for each object in the catalogue of 
\cite{01CAY_EA}, and demonstrate the underestimate with no ambiguity.

Together with errors on effective temperatures, the 2 biases  described here are sufficient 
to stretch the metallicities and produce the salient clump in the 2 AMDs of Nordstr\"om et al. (2004)
and Feltzing et al. (2001). After having corrected these defects, it is shown in section 4 that 
stars at [Fe/H]$<$ 3 Gyrs have a metallicity in continuity with older stars, and do not form a specific pattern.

\subsubsection{Young metal-poor stars}

\begin{figure}
\includegraphics[width=8cm]{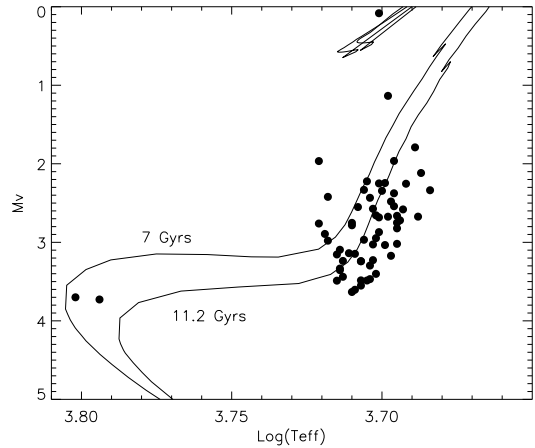}
\caption{Stars in GCS catalogue with age $<$9 Gyrs, [Fe/H]$<$-0.4 and B-V$>$0.75. 
The 2 isochrones are from Girardi et al. (2000) and have ages of 7 and 11.2 Gyrs, Z=0.004.
In Nordstr\"om et al., models at this metallicity have been shifted by 0.01-0.015 dex in Log(T$_{\rm eff}$)
because of problems in main sequence effective temperatures.  However, the visible consequence
is that stars at the base of the red giant branch will be more coincident with a 7 Gyrs or younger isochrone. 
Comparison of these stars with the (M$_v$,B-V) fiducial sequence of 47 Tuc shows they are genuinely old.}
\label{oldyoung}
\end{figure}

Inspection of Fig \ref{fehage}(b) shows that, at  medium to low 
metallicities in the GCS catalogue, young stars are more numerous than old stars. 
There are 522 stars with [Fe/H]$<$-0.4 and age $>$ 6 Gyrs,
but 829 with age smaller than this limit. At [Fe/H]$<$-0.6 (fully in the thick disc regime), 
'old' stars (as just defined) are the majority, but there are however 45 \% of 'young' objects. 
If the limit is shifted to 9 Gyrs, this percentage rises to 72\%.
Although these stars represent a minor subset of the whole catalogue, they are essential
to characterise the age-metallicity distribution of Nordstr\"om et al. (2004), 
since they give the impression that old stars are not particularly deficient stars. 
The characteristics of these stars show that they can be classified in 2 categories. 
The first category is made of the youngest stars (age $<$ 3 Gyrs) which mostly have 
B-V$<$0.5, and whose metallicity is underestimated.  These were studied in the previous section. 
Note that, among this subsample, a few stars at B-V$<$0.5 are probably genuinely young, 
with metallicities that are truly deficient relatively to their age. 
This is confirmed for objects which have spectroscopic metallicities such as HIP 116082, 
32851, 47048 or 83243. These are discussed further below.
The second category is the group of cold objects (B-V $>$ 0.75). There are 73 such objects
with measured ages and  [Fe/H]$<$-0.4 dex in the GCS catalogue, 
51 with age younger than 7 Gyrs. These are mostly stars at the 
beginning of the red-giant branch. For this peculiar subsample, the few objects in common with the catalogue of Cayrel et al. (2001)
don't show significant systematics in effective temperature or metallicity. 
When compared with the isochrone of Girardi et al. (2000) 
(used by  Nordstr\"om et al. (2004)), it can be verified that these objects are correctly 
fitted by old isochrones (10-12 Gyrs) with intermediate metallicities, see Fig \ref{oldyoung}, although they
are several Gyrs younger in Nordstr\"om et al. (2004). However, 
in this metallicity range, Nordstr\"om et al. have shifted stellar models 
by $\approx$ -0.011 dex  in Log T$_{\rm eff}$, because of a known discrepancy between model 
and {\it main sequence} data in this metallicity range (Lebreton 2000). There is no  evidence 
that this discrepancy also 
concerns the base of  the red giant branch. At M$_v$ = 2.8-3.0, typical 
of these objects, the difference in Log(T$_{\rm eff}$)  between a 7 and 11 Gyrs 
isochrone is 0.01 dex (see Fig. \ref{oldyoung}), the amount of the temperature correction applied by \cite{04NOR_EA}.
Shifting models to lower temperatures is equivalent to shifting observed stars towards higher 
temperatures, producing the effect that is seen on Fig. \ref{allbias}, or creating young, metal-poor objects. 
Therefore, it is not surprising that the majority of these stars are found at rather 
young ages in the GCS catalogue,  while the old metal-poor region is correlatively depleted. 
Comparison between the position of the fiducial sequence of 47 Tuc and these stars in the HR diagram 
confirms that they are old objects.

\subsubsection{The scatter in the age-metallicity relation}

Under- and over-estimated metallicities in Nordstr\"om et al. (2004) and Feltzing et al. (2001) 
(see Fig \ref{biasedfeh})  stretch the AMD by at least $\pm$0.1-0.2 dex at ages $<$ 3 Gyrs.
Fig. 6 (simulated biases) shows that, over the same age interval, combining these two biases 
scatters the metallicities from about -0.4 to +0.4 dex when effective temperatures are overestimated, 
and from -0.6 to +0.4 dex when they are underestimated, even though we started
from a monotonic age-metallicity relation with zero dispersion.
We show in the next section that, with revised ages, the dispersion at $<$ 3 Gyrs is most probably 
about 0.1 dex, which can mostly be attributed to the observational scatter
in photometric metallicities.

Simulated biases show that underestimated effective temperatures uniformly 
populate the AMD between -0.6$<$ [Fe/H] $<$ 0.0 dex at all ages, increasing the rms dispersion 
to 0.2 dex  for the oldest stars. 
Given the fact that our simulated biases are oversimplistic, that many effects have not been 
taken into account (we assumed no intrinsic cosmic scatter), that our model age-metallicity relation is 
probably overestimating  the change of metallicity, we see it as a normal consequence that a uniform scatter
in the AMD has been found from photometric surveys, even though the real intrinsic cosmic scatter
could be as small as the one measured in the local ISM, on meteoritic pre-solar dust grains, or  
on element ratios (that is $<$ 0.04 dex). 

In any case, we emphasize that the (B-V, [Fe/H]) plots (Fig. \ref{bvfehgen}) show unambiguously that
the cosmic scatter increases towards redder, hence older, stars. This, and the above arguments, 
strongly suggest that the large scatter in metallicity {\it at all ages}\footnote{We suggest hereafter, section 4,
that the distribution of metallicities at a given age is probably the combination of 2 components, one
with a small dispersion, the other, with greater dispersion, made of stars that migrated to the 
solar neighbourhood; the importance of this component should increase with age.}
found in Nordstr\"om et al. (2004) 
and Feltzing et al. (2001),  is not real.

\section{A revised age-metallicity distribution}

In order to illustrate the sensitivity of the final AMD
to initial inputs (metallicities, temperatures), we now re-calculate 
the age-metallicity distribution with modified effective
temperatures and different metallicities. 
Following our comments in section 3.2.1, we first correct the effective temperatures 
of Nordstr\"om et al. according to the correction calculated on Fig. \ref{diBe}. 
Although that will reduce or eliminate the systematic shift 
in temperature, the observational dispersion will (obviously) remain.
Reducing the random dispersion on effective temperatures would imply 
different strategies and data and is beyond the scope of the present paper.

\begin{figure}
\includegraphics[width=9cm]{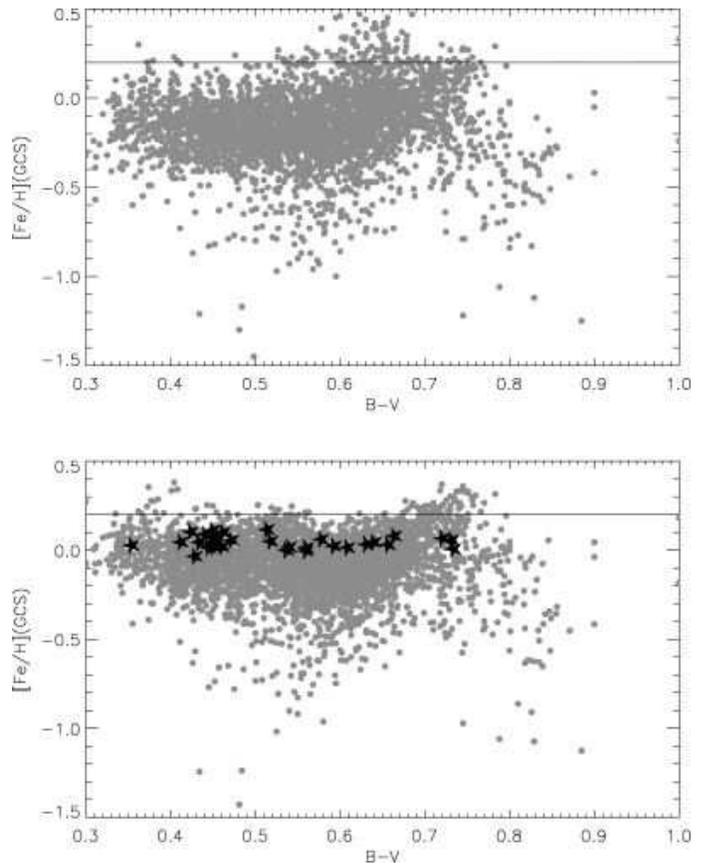}
\caption{(a) (B-V,[Fe/H]) for a subsample of the GCS catalogue, selected as described in the text. 
(b) (B-V,[Fe/H]) for these same stars with metallicity derived as described in the text.
}
\label{revisedbvfeh}
\end{figure}

Correcting the metallicities given in the two cited studies is difficult
because of the complex intrication of the different biases with B-V colour 
and Str\"omgren indices. The simplest procedure is to adopt another similar
metallicity calibration, which shows significant improvement over those 
used by Nordstr\"om et al. or Feltzing et al.
We have used the calibration given by Ram\'irez \& M\'elendez (2005a) which 
is of similar form as the one by Schuster \& Nissen (1989), but seems to be
well behaved in a (B-V,[Fe/H]) plot (Fig.\ref{revisedbvfeh}). Note however that a small colour term in the sense
that [Fe/H] decreases with B-V (seen in the Hyades stars as well as in field stars) 
has been corrected as follows: [Fe/H]=[Fe/H]$_{\rm RM}$-0.5*(B-V)+0.35 for stars
bluer than B-V=0.7. 
The calibration has been applied 
to the stars that have b-y, m$_1$, c$_1$ indices in \cite{93OLS} and \cite{94OLS2},
 and a parallax in the Hipparcos catalogue with relative error less than
10\% (or distance less than $\sim$ 100pc), in order to minimise reddening effects. 
\cite{83OLS} has not been used,
because comparison with the two other datasets showed significant inconsistencies which
contribute to the dispersion in metallicities.
No correction for reddening was adopted. 
In case a star was found in the 2 catalogues, the mean of the 2 determinations was 
adopted. 
This selection yields 4469 stars for which (B-V, [Fe/H]) is plotted in Fig. \ref{revisedbvfeh}(a), with
metallicities from the GCS catalogue, while plot (b) shows the new metallicities.
We note in passing that the general form of these two plots is similar to that of the Geneva sample (Fig. 9a), 
and different from the Fig. \ref{fehage}(a).
The main differences between Fig. \ref{revisedbvfeh}(a) and (b) are (1) the absence of the metallicity 
uprise between B-V=0.6 and 0.7 and [Fe/H]$>$+0.2 dex in (b), which was due to overestimated values in the GCS catalogue,
(2) the substantially lower metallicities at B-V$<$0.6, also present in the calibration of Ramirez \& M\'elendez (2005a), 
and that was corrected as explained previously. 

\begin{figure}
\includegraphics[width=9cm]{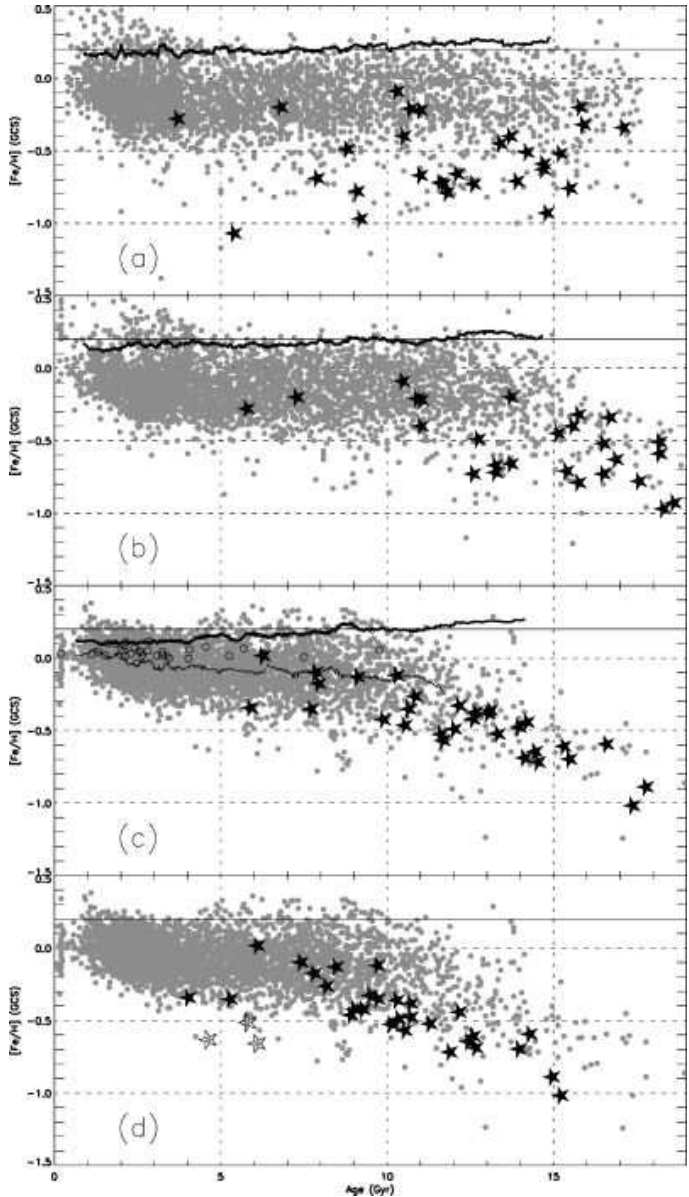}
\caption{AMD for various age and metallicity determinations. The horizontal line at [Fe/H]=0.2 dex is 
given to ease the comparison. The thick curve on the first 3 plots is the dispersion in metallicity
calculated over 300 points. The solid stars in all plots are thick disc stars according to their kinematics
(see text).
{\bf (a)} The AMD  according to the subsample of the GCS catalogue, for stars selected as detailed
in the text. Ages and metallicities are all from the CGS catalogue.
{\bf(b)} The AMD for the same stars, with the atmospheric parameters from the GCS catalogue, but with ages derived 
from the Yi et al. (2003) isochrones. The $\alpha$-element content was set to 0 for all stars. 
{\bf(c)} AMD for the same sample, but derived from the new [Fe/H] (see text),  effective temperatures 
from the GCS corrected as described in the text, and using
the isochrones of Yi et al. (2003). The $\alpha$-element content was set to 0 for all stars.
The open circles  are the new ages derived for the Hyades stars.
The thin lower curve is the mean metallicity calculated over 100 points. 
{\bf(d)} Same as (c), except that the ages of thick disc stars (solid stars) have been derived taking into account the 
$\alpha$-element content according to the value listed in Soubiran \& Girard (2005). 
Open stars are objects with [Fe/H]$<$-0.4 from Soubiran \& Girard (2005) flagged as disc stars according to their kinematics
and $\alpha$-element content.
}
\label{revisedamr}
\end{figure}

About 3650 objects in this sample have an age in the GCS catalogue. 
The AMD from the GCS catalogue for these stars is given on Fig \ref{revisedamr}(a), and although it is a subsample,  
it retains all the main characteristics of the AMD of the GCS catalogue.
Fig. \ref{revisedamr}(b) has been obtained from the same parameters (Teff, [Fe/H]), but with
ages derived with the method employed in our study and the isochrones of \cite{03YI_EA}.
It illustrates that changing the dating method and isochrones doesn't change
significantly the shape of the AMD, although it must be noted that most of the young
metal-poor stars in the GCS catalogue selection are shifted towards older ages. This is 
illustrated in particular by the thick disc stars (star symbols, see below for a discussion of these objects). 
Fig. \ref{revisedamr}(c) shows the AMD for the same stars with age redetermined using the  \cite{03YI_EA}
isochrones and corrected effective temperatures and metallicities.
The circle symbols show the new ages and metallicities for the Hyades stars. 
Compared to the GCS catalogue parameters for the Hyades, metallicities are nearer to the expected 
metallicity of this cluster, and are independent of B-V (within the uncertainties).
Obviously, the derivation of ages is still not optimal and still
affected by significant errors. This is not surprising since we have not optimized 
the derivation of effective temperatures nor metallicities. 

There is however a clear indication that the newly derived ages and metallities are superior to those from the GCS catalogue.
This is seen by comparing the dispersion in metallicity as a function of B-V (Fig. \ref{revisedbvfeh}) and age 
(Fig. \ref{revisedamr}) for the 2 datasets. 
If there is an (even loose) relation between age and metallicity, it is expected that 
the dispersion in metallicity be 
lower when correlated with age, because B-V
is only representing a mean age sequence. However, in the case of the GCS catalogue, the dispersion is 
larger at ages $<$ 4 Gyrs (0.16-0.19 dex, Fig. \ref{revisedamr}(a)) than in B-V (for B-V$<$0.47, Fig. \ref{revisedbvfeh}(a)) (0.14-0.16 dex), whereas in our case, the dispersions
are 0.10-0.11 (Fig. \ref{revisedamr}(c)) and 0.13-0.16 dex (Fig. \ref{revisedbvfeh}(b)) over the same intervals. 

Interestingly, the AMD we obtain with our new age and metallicity determinations is quite different 
from that of Nordstr\"om et al. (2004) and Feltzing et al. (2001), and we now give a detailed account 
of the differences.

(a) Three of the characteristics previously discussed (the clump; young, intermediate metallicity stars;
decrease of metallicity between 10 and 5 Gyrs) are now absent. 
The overall dispersion varies from below 0.11 dex at age $<$ 3 Gyrs to more than 0.22 dex
at ages $>$ 10 Gyrs. There is a significant difference with the dispersion measured on the data
of Nordstr\"om et al. (2004) on Fig. 13a, which is about 0.17 dex at ages $<$ 3 Gyrs 
 Interestingly a dispersion of 0.1 dex is about what is expected from measurement uncertainty, which means
that intrinsic scatter at ages $<$ 3 Gyrs is compatible, within the uncertainties, 
with the one measured on abundances in
the interstellar medium and isotopic ratios of meteoritic presolar dust grains ($\approx$0.04 dex, 
Cartledge et al. 2006, Nittler 2005). 

(b)  The increase in the metallicity range towards super metal-rich objects at ages greater than 4-5 Gyrs
corresponds to a similar increase in the (B-V,[Fe/H]) plot at B-V$>$0.6 (Fig. \ref{revisedbvfeh},
the same is also observed with Geneva and spectroscopic metallicities, Fig. \ref{bvfehgen}).
This feature, which is seen neither in Nordstr\"om et al. (2004) nor 
Feltzing et al. (2001),  shows the inclusion in the solar neighbourhood of objects that are
gradually  older and more metal-rich. The (B-V,[Fe/H]) plot of the Geneva sample shows 
that these objects reach [Fe/H]=+0.5 dex at 5 Gyrs. The extent of these features at 
B-V$>$ 0.8 in this last sample suggest that the absence of stars older than 9-10 Gyrs on Fig. \ref{revisedamr}(c)
is probably a selection effect of the Str\"omgren sample (the turn-off colour of a 10 Gyr isochrone
at [Fe/H]=0.3 dex is B-V=0.75).

The sample of Geneva metallicities (B-V,[Fe/H]) in Fig.\ref{bvfehgen}(a) shows that local young 
objects (B-V$<$0.4) reach a maximum metallicity similar to that of the Hyades cluster 
([Fe/H]=0.10-0.15 dex).
If the origin of the stars more metal rich than this limit is radial migration (Grenon 1999), and given an estimate 
of the radial metallicity gradient, it could in principle  be feasible to measure an upper limit to
the rate of radial migration. 
However, the radial gradient is uncertain, with values between 0.04 to 0.1 dex/kpc, implying  
a maximum rate of migration loosely constrained between 1.75 and 0.7 kpc/Gyr.

(c) There is a similar extension of the lower metallicity interval that intervenes at ages $>$ 4 Gyrs.
For only 3 of these objects, an $\alpha$ abundance ratio could be found (open star symbols in 
Fig.\ref{revisedamr}(d)), showing with no ambiguity that these are bona fide thin disc stars.
Pont \& Eyer (2004) have commented upon young, metal-poor stars in Edvardsson et al. (1993), explaining
that these objects were unlikely to be truly young, since none
had a measured mass (from fitting the HR position with an isochrone) greater than 1.1 M$_{\odot}$. 
We emphasize that unlike the stars studied by Pont \& Eyer (2004), all these three objects have 
an estimated mass higher than this limit. Also, given their scarcity (7 objects with [Fe/H]$<$-0.5 dex and 
age $<$ 7 Gyrs out of 3650 stars in our sample) , it is unsurprising that the Edvardsson et al. (1993) 
sample contains none of these objects. 
We note that these metal-poor 'young' candidates seems to appear at a similar age as metal-rich objects,
although statistically more significant samples are needed, and this can be 
attributable to a similar cause (radial migration), but from the outer disc.

(d)  The median value calculated in (overlapping) subsamples of 200 points between 10 Gyrs 
and the youngest stars shows that the increase of metallicity between the 2 limits is of 
the order of  -0.15 dex, with a  small upturn at age $<$ 3 Gyrs.
There is a change of slope when shifting  from the thin to the thick disc, for which the metallicity increases by about 0.5 dex 
within 5 Gyrs.

(e) As a final test to the effect of our corrections on the atmospheric parameters, we selected stars 
that are known to belong to the thick  disc according to their kinematics. We have cross-correlated our sample 
with the compilation of stars published by Soubiran \& Girard (2005), which yielded 29 stars.  
These 29 stars are shown on all four plots of Fig. \ref{revisedamr} as solid star symbols.
The most remarkable result is seen when comparing the different age distributions. In the first
case (plot (a), GCS catalogue), no particular trend is seen, and the spread of points illustrates 
our comments on the various biases (see the previous section).
Part of this spread is reduced with the ages derived from the isochrones of Yi, Kim \& Demarque (2003), 
due to the fact that no correction to the effective temperature scale of the models has been applied, 
hence we avoid  the bias detailed in section 3.3.3, which generates young metal-poor stars. 
The new ages calculated with the corrected atmospheric parameters (Fig. \ref{revisedamr}(c)) 
show a clear, distinct trend of increasing metallicity with decreasing age, and confirms the 
existence of an age-metallicity relation within the thick disc proposed by Bensby, Feltzing \& 
Lundstr{\" o}m (2004). In this case, 
we have neglected the $\alpha$-element content and set [$\alpha$/Fe]=0.0 dex.
Finally, in Fig. \ref{revisedamr}(d), we use the value of  $\alpha$-element content 
listed in Soubiran \& Girard (2005) for each star to compute the isochrones and derived the ages. 
The correlation obtained is even better than in the previous plot (c).

\section{Discussion} 

We reviewed the metallicity distribution of stars in the solar neighbourhood 
published in the recent years  and find that  recent spectroscopic datasets show good 
agreement with our own finding (Haywood 2001),  with a peak at [Fe/H]=-0.05. 
Using scale height corrections based on a $\sigma_W$-[Fe/H] relation
derived from a sample of solar neighbourhood stars,  we arrive at the conclusion that
there is no deficit of metal poor stars relative to the  closed box model.
The low metallicity part ([Fe/H]$<$-0.5 dex) of our relation being dominated by thick 
disc stars, this conclusion is valid only if the thick disc stars are genuine disc stars, 
not of extragalactic origin.  It is also in agreement with expectations from simple
arguments of Galactic structure. 
Standard chemical evolution models usually acknowledge the existence of a thick disc 
as a genuine Galactic population, but its contribution to the local solar radius 
metallicity distribution is often underevaluated. On the other hand, looking at the 
metallicity distributions recently published, we find the contribution of stars 
having the metallicity of the thick disc to be generally incompatible with simple
Galactic structure constraints.

Despite the recent detailed analysis of large photometric Str\"omgren data sets, it appears that published
AMD  are essentially reflecting noise in the determination 
of atmospheric parameters and derived ages. 
Significant improvement should come from reducing systematic
biases in the determination of metallicity and effective temperatures from photometric indices.
A first step in this direction is made here, and shows a limited increase in the mean metallicity 
of the thin disc, superposed on another, steeper relation in the thick 
disc. This scheme confirms that the disc has endured most of its chemical evolution during the thick disc phase, 
and has remained (chemically) unchanged since then.

In the last 3 Gyrs, the measured dispersion in metallicity is 0.10-0.11 dex, compatible with error
measurements, and implying a small scatter of the young disc population.  
After 3 Gyrs, the progressive appearance of  a metal rich population reaching [Fe/H]$\approx$+0.5 dex 
at ages greater than 5 Gyrs is established. It is suggested that these super metal-rich stars are the contaminant objects resulting
from radial migration  from the inner disc. A similar spread at lower metallicity is seen at approximately 
the same age, giving stars reaching [Fe/H]$\approx$-0.7 at 5 Gyrs, attributable to an equivalent migration from the 
outer disc. These characteristics are consistent with the (B-V, [Fe/H]) distribution presented in section 3, 
which, although independent of any age determination, is a source of information on the AMD. 
Note also that the metallicity scale of the  (B-V, [Fe/H]) distribution is independent of the Str\"omgren 
metallicity scale used in the derivation of ages.
If this general picture is correct, we expect that the metallicity distribution at age greater than 3 Gyrs is 
a superposition of a narrow (about 0.05 dex or less) distribution of stars born at solar radius and a broader 
distribution of stars that haved moved to the solar radius. We note that the metallicity
distribution of Fuhrmann (2004, figs. 49 and 50) presents such characteristics, with clearly a 
superposition of 2 gaussians with dispersions of about 0.05 and 0.25 dex centred on solar metallicity.

Altogether, these elements point to a picture  where chemical and dynamical evolution lead to 
structured, if complex,  patterns of age and metallicity, but not a general, structureless, dispersion.
It is apparent that, in the level of abundance reached by the Galaxy prior to the thin disc formation
as well as in the shape of the local metallicity distribution, the thick disc must have played a central role.
This is now discussed.

\subsection{The thick disc : Is there a continuity argument ? }

Is there a continuity argument to invoke when discussing the origin of the thick disc ? 
Inspection of  $\alpha$ elements ratios $vs$ metallicity plots (see Pritzl, Venn \& Irwin (2005), 
Reddy, Lambert, \& Allende Prieto (2006) for example)  suggests the following : 
(1)  The halo presents no evidence of having been enriched by SNeIa
(even when allowing for a generous thresold in galactocentric rotational velocity at about +100-120 km/s, see Fig. 2 in
Pritzl et al, 2005). There have been suggestions that low [$\alpha$/Fe] implied contamination by SNeIa. 
However, most recent studies (Arnone et al. (2005) and references therein) indicate that  halo stars seems to have a higher level of homogeneity
in element ratios [X/Fe] than previously thought ($<$ 0.06 dex), with levels of [$\alpha$/Fe]
compatible with no contamination by SNIa.
(2)  while there is a clearly visible increase of the [$\alpha$/Fe] ratio towards low metallicity stars
of the thin disc ([$\alpha$/Fe]=0.1--0.2 dex at about [Fe/H]$\approx$-0.5 -0.7 dex, see Reddy et al. (2006), Fig. 12), 
there is no evidence that the thin disc has escaped contamination by SNIa.
(3) in the solar neighbourhood at least, the thick disc seems to be the {\it only} population that shows 
both enrichment phases by SNeII and SNeII+SNeIa, 
suggesting that the thick disc could be a transition population between the halo and
thin disc. The suggestion is made stronger if one realises that an extragalactic thick disc would require that 
an accreted satellite  have just the right ($\alpha$/Fe, Fe/H) pattern to meet the halo on one side and 
the thin disc on the other.
As a matter of fact, there is no detected chemical or kinematical discontinuity between the 'rotating halo'  
and thick disc (see Gratton et al. 2003) so far. 
If the small dispersion of abundance ratios measured on halo stars is confirmed, then the continuity between 
enrichment levels in the different elements between the halo and the thick disc should 
become a stringent constraint for the origin of this last population. 
For example, Nissen et al. (1994) measured 
[Mg/Fe]=0.41 with a rms dispersion of $<$0.06 dex on halo stars with -3.2$<$[Fe/H]$<$-1.8, while 
Reddy et al. (2006 and references therein) measured thick disc stars to have a similar rms dispersion 
and [Mg/Fe] comparable with that of Nissen et al. (1994) 
at [Fe/H]$<$-0.8dex. Confirmation on a larger scale is needed, but it suggests that the 'rotating halo' and the
thick disc may be the same population. 

There is however an observed discontinuity between the thin and thick discs, with the  ($\alpha$/Fe, Fe/H) sequence of the two
being almost parallel.
It has been proposed that since the most metal-rich stars in the thick disc have a higher metallicity
than the most metal-poor stars in the thin disc, there has been a dilution of metals by an infall
episode in the temporal  gap between the two populations (Bensby et al. 2005; Reddy et al. 2006).
The AMD of the previous section shows that metal-poor disc stars are not specifically old for their metallicity, 
a characteristic that is not expected if it is assumed that the first thin disc stars had 
formed after the infall episode at [Fe/H]$\approx$-0.6 dex. Moreover, the AMD of Fig. 13 shows that the 
mean metallicity of old disc stars is nearer to [Fe/H]=-0.2 dex than -0.6 dex, a fact which does not
fit well with the dilution scenario above.
A perhaps more convenient explanation is that metal-poor thin disc stars found in local samples have been moved to 
the solar neighbourhood by radial migration from outside the solar  circle. 
This scenario, suggested by our conclusions that the AMD has been shaped by radial migration, 
is also well in agreement with
the results presented by Carney et al. (2005) and Yong, Carney \& de Almeida (2005).
These authors present evidences, from open clusters, field giants and cepheids, that the mean disc metallicity
decreases to (-0.4, -0.6) dex at R$_{GC}$=10-11 kpc. An even more interesting clue found by these authors is
that the [$\alpha$/Fe] ratios of these stars reach about +0.1 to +0.2 dex. 
This is just the characteristics of the most metal-poor thin disc stars in the solar neighbourhood sample, which 
also show this slight increase in  [$\alpha$/Fe].
Finally, we note that Carney et al. (2005)  and Yong et al. (2005) find that the gradient in metallicity 
and [$\alpha$/Fe] flattens beyond  R$_{GC}$=10-11 kpc, which may explain why the metal-poor thin disc stars
found in local samples are limited to about  [Fe/H]$\approx$-0.6 dex.
Taking these considerations  into account, it might then be expected that the thin disc starts forming its
stars (at solar galactocentric radius) not at -0.6 dex, but at a higher metallicity which Fig. 13 suggest
might be around -0.2 dex, or above. This is also consistent with the ([$\alpha$/Fe], [Fe/H]) plot in 
Reddy et al. (2006), which, showing excellently the separation between the thin and thick discs, also
shows that the tip of the metal rich thick disc is around (-0.3,-0.2) dex.

This leaves the possibility that there is a gap in  [$\alpha$/Fe] between the thin disc and the 
thick disc (Pagel 2001)
of the order of +0.05 dex, consistent with a temporal gap between the two populations. 
A jump in  [$\alpha$/Fe] with no significant variation of the general metal content ($<$0.1 dex) does not 
require exchange of gas (inflow or outflow) but simply that star formation ceased for a period of time
sufficiently long that the ISM was enriched in iron by SNeIa for [$\alpha$/Fe] to decrease by 
0.05 dex.

\subsection{Life without the 'G-dwarf problem'}

In view of what has just been said about the thick disc, 
how do we interpret the fit between the dwarf metallicity distribution and the closed-box model ?
We cannot eliminate the option that the thick disc is an accreted population, since there is by now 
no clear-cut evidence against this solution. In this eventuality however, we note that the metallicity
distribution would then have to be limited to 'pure' thin disc, solar radius stars. 
Limiting the sample to objects born within a restricted range around this radius would 
probably make the metallicity much thinner than the one derived in Haywood (2001) - probably less
than 0.1 dex dispersion. 
Casuso \& Beckman (2004) have explicitely considered the thick disc should not be part of the metallicity
distribution,  although not formally considering the thick disc as a population of extragalactic 
origin.
Their conclusion that infall should be an increasing function of time, is not surprising.

In the case that the thick disc is a genuine Galactic population, the disc (thin + thick) 
metallicity distribution is not in marked disagreement with the closed-box model.
The problem is reminiscent of the discussion about the MDF (metallicity distribution function) of the halo. According to predictions of hierarchical
clustering in $\Lambda$CDM cosmologies, the stellar halo is supposed to be an archetypal open system, formed from
many independent units. However, the observed MDF of the halo is shown to be 
fitted reasonably well by a simple box model distribution (Oey 2003) over a large range of metallicities.
More generally, it is apparent that the closed-box model gives a good fit to the
MDF of spheroids for most of the metallicity interval. There is often a problem in fitting the metal poor tail of
these distributions, but this concerns a small fraction of the stars, and 
is not comparable with the local disc 'G-dwarf problem'.
The question however remains : why does the closed box model provides an honest fit to
these systems ? 
The question for the disc MDF is more or less the same. The exact profile of the low metallicity tail 
of the thick disc is unknown, but globally, the simple model provides a fair fit to the observed distribution.
Because discs are now conceived as open systems that build up continuously from accreted gas, this result 
is not expected. We emphasized that strictly speaking, our result only means that the allegated 
mismatch between the observed MDF and the simple model cannot be used as an argument for
infall models. It does not imply infall is not a crucial ingredient in models, 
although it strongly suggests that standard infall models should be revised.
We notice that first results from chemo-dynamical models 
(Brook et al. 2005) are encouraging. 
 In particular, Brook et al. (2005) have come quite close to reproducing the characteristics of 
the distinct ($\alpha$-element) chemical signatures of the thin and thick discs, with the 
thick disc formed during a merging phase of  gas-rich  'building block', while the thin disc
forms inside-out, at a more quiescent epoch, from continuously accreting gas.
The interesting point is given by inspection of the metallicity distribution of the (thin + thick) disc 
generated by their model (Martel et al., 2005, their Fig. 3), which shows it 
is quite similar to a closed box model distribution. 
This suggests that although the local data are not conclusive indication for prolonged infall, as classically
proposed, a metallicity distribution with the simple model characteristics is neither an 
indication of a closed system\footnote{Ironically though, 
it may be noted that at least 2 hypotheses  of the simple  model are little disputed,
and one fact is hard to prove wrong : that the IMF is constant with time, 
the interstellar medium is well mixed, and, with the present results, that the solar neighbourhood 
metallicity ditribution is like a closed box model distribution.}.
Said differently,  the simple model may have lost its paradigmatic strength.

\section*{Acknowledgments}
I would like to thank the referee for the very helpful comments and 
suggestions which much improved the first version of this paper.
  \label{lastpage}
 
\bsp

\end{document}